\documentclass[aps, prx, superscriptaddress,reprint]{revtex4-2}

\usepackage{braket,amsmath,amssymb,graphicx,comment,xcolor,float}
\usepackage[T1]{fontenc}
\usepackage{times}
\usepackage{xcolor}
\usepackage{hyperref}
\usepackage{siunitx}
\usepackage{ulem}

\newcommand{\refsub}[2]{\hyperref[#1]{\ref*{#1}#2}}

\usepackage{overpic}

\begin{document}

\title{Control and Entanglement of Individual Rydberg Atoms Near a Nanoscale Device} 

\author{Paloma L. Ocola}
\altaffiliation{These authors contributed equally to this work}
\affiliation{Department of Physics, Harvard University, Cambridge, MA 02138, USA}
\author{Ivana Dimitrova}
\altaffiliation{These authors contributed equally to this work}
\affiliation{Department of Physics, Harvard University, Cambridge, MA 02138, USA}
\author{Brandon Grinkemeyer}
\altaffiliation{These authors contributed equally to this work}
\affiliation{Department of Physics, Harvard University, Cambridge, MA 02138, USA}
\author{Elmer Guardado-Sanchez}
\altaffiliation{These authors contributed equally to this work}
\affiliation{Department of Physics, Harvard University, Cambridge, MA 02138, USA}
\author{Tamara \DJ{}or\dj{}evi\ifmmode \acute{c}\else \'{c}\fi{}}
\affiliation{Department of Physics, Harvard University, Cambridge, MA 02138, USA}
\author{Polnop Samutpraphoot}
\affiliation{Department of Physics, Harvard University, Cambridge, MA 02138, USA}
\author{Vladan Vuleti\ifmmode \acute{c}\else \'{c}\fi{}}
\affiliation{Massachusetts Institute of Technology, Cambridge, MA 02139, USA}
\author{Mikhail D. Lukin}
\email[To whom correspondence should be addressed; E-mail: ]{lukin@physics.harvard.edu}
\affiliation{Department of Physics, Harvard University, Cambridge, MA 02138, USA}


\begin{abstract}
Rydberg atom arrays constitute a promising quantum information platform, where control over several hundred qubits has been demonstrated. Further scaling could significantly benefit from coupling to integrated optical or electronic devices, enabling quantum networking and new control tools, but this integration is challenging due to Rydberg sensitivity to the electric field noise from surfaces. We demonstrate that Rydberg coherence and two-atom entanglement can be generated and maintained at distances $\sim100\mu$\SI{}{m} from a nanoscale dielectric device. Using coherent manipulation of individual qubits and entanglement-assisted sensing, we map the spatio-temporal properties of the electric field environment, enabling its control and the integration of Rydberg arrays with micro- and nanoscale devices.
\end{abstract}

\date{\today}
\maketitle

Significant progress is currently being made in developing quantum information processors, promising to tackle computationally difficult problems. However, further increasing the computational power may require connecting multiple processors via quantum interconnects~\cite{monroe13}. Rydberg atom arrays have recently emerged as a leading platform for quantum simulations and quantum information processing, 
where entangled states of matter,  tests of quantum algorithms and quantum error correction are currently being explored with hundreds of qubits~\cite{saffman2010quantum,BrowaeysLahaye20,ebadi22}.
While scaling to many thousands of controlled qubits appears feasible~\cite{saffman2022}, significant advances can be achieved by coupling Rydberg arrays to optical, microwave and electronic devices. Integration with these devices could enable quantum networking via optical photons \cite{reiserer15,northup14,huie2021multiplexed,young22} as well as novel coupling and control techniques via microwave photons~\cite{sorensen2004}. In practice, however, Rydberg qubits experience decoherence near surfaces caused by fluctuating charges. While these effects have been studied near various types of dielectric \cite{spreeuw18,shaffer16,rajasree20,kubler2010coherent,epple14,carter2013}
and superconducting \cite{hermann2014,thiele2015,gard2017,kaiser22,kumar2022}
surfaces, Rydberg atom integration with such devices remains challenging.

\begin{figure}[h!]
\centering
\includegraphics[width=0.9\columnwidth]{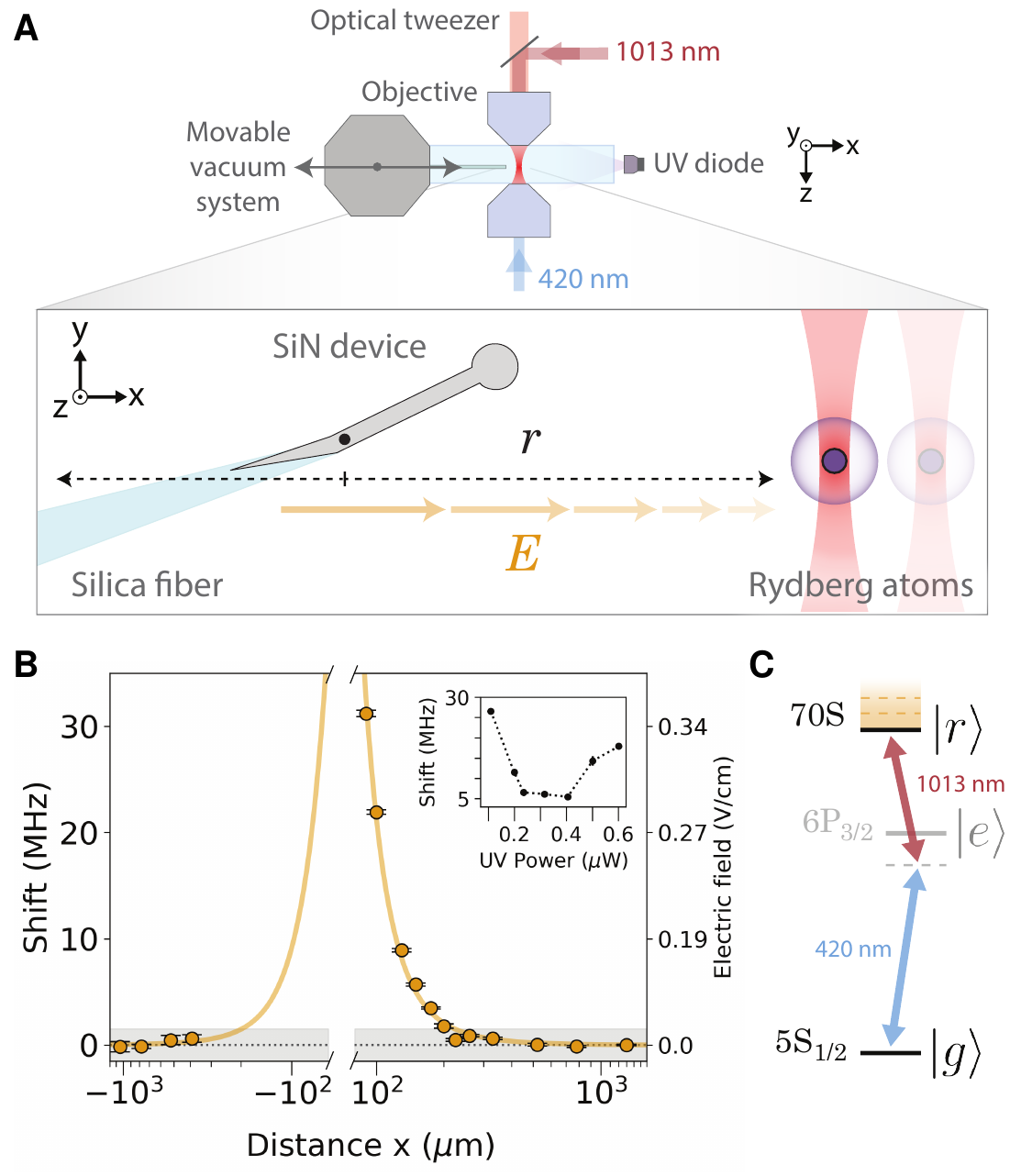}
\caption{\textbf{Rydberg atoms near a nanoscale photonic device.} \textbf{(A)} A nanoscale SiN device attached to a tapered fiber in a vacuum chamber is moved relative to a stationary optical tweezer. \textbf{(B)} Measured spectral shift of the $\ket{70S}$ Rydberg state as a function of distance from the device fitted to a point-charge of \SI{126(11)}{e} located on the device (solid line) that exceeds the Rabi-broadened linewidth (shaded gray region) for distances $<\SI{200}{\um}$. (Inset) Spectral shift dependence on UV power at \SI{160}{\um}. \textbf{(C)} \SI{420}{\nm} and \SI{1013}{\nm} light used to excite to the $\ket{70S}$ Rydberg state which is shifted by electric fields.}
\label{fig:fig1}
\end{figure}

In this Report, we explore the coherence properties of Rydberg atom qubits and entangled states in close proximity to a silicon nitride (SiN) nanophotonic crystal cavity, used previously to achieve atom-photon and transportable atom-atom entanglement~\cite{tiecke14,dordevic21}. Remarkably, the electric field from this nanoscale device resembles a point-charge of $\sim 200$ single electron charges $(\SI{}{e})$ with quasi-static fluctuations, enabling coherent control via decoupling pulse-sequences at distances as close as \SI{100}{\um} from the device. Moreover, we demonstrate that certain entangled states are relatively insensitive to charge noise, allowing us to perform an entanglement-assisted measurement of the electric field environment and study its control. Together with the recently demonstrated coherent transport of ground state atoms \cite{bluvstein22,dordevic21}, these observations open the door for integration of Rydberg arrays with complex optical, microwave and electronic devices.

In our experiments, individual Rubidium-87 atoms are trapped in optical tweezers and placed at varying distances from a nanoscale device. The device is suspended on a tapered silica fiber connected to a translation stage (Fig.~\refsub{fig:fig1}{A}), allowing it to be positioned $90-2600\SI{}{\um}$ from the trapped atoms. Each atom is prepared in $\ket{g}=\ket{5S_{1/2}, F=2, m_F=2}$ and excited to the $\ket{r} = \ket{70S, J=1/2, m_J=1/2}$ Rydberg state via a two-photon transition  (Fig.~\refsub{fig:fig1}{C}). The excitation beams are focused onto the atom while the tweezer light is off and Rydberg population is detected by atom loss.
The Rydberg state experiences a Stark shift $\Delta \nu = \frac{1}{2} \alpha |\mathbf{E}|^2$ in an electric field $\mathbf{E}$ with polarizability $\alpha = \SI{534}{\mega\Hz/(\V/\cm)^2}$~\cite{osullivan1985}. Its spectral shift is measured as a function of distance from the device (Fig.~\refsub{fig:fig1}{B}), exceeding the Rabi-broadened linewidth of \SI{3}{\MHz} for $|x|<\SI{250}{\um}$. The shift follows the electric field scaling of a point-charge ($r^{-4}$) located on the \SI{31.5}{\um}-long device (Fig.~\refsub{fig:fig1}{C}). Assuming the charge is at the center of the device gives an estimated charge $q=$\SI{126(11)}{e}~\cite{supp}. At each distance the shift is minimized by illuminating the experimental setup with a UV diode and optimizing its power (Fig.~\refsub{fig:fig1}{B, inset})~\cite{supp}. Furthermore, the spectral shift is stable only with a relatively constant rate of Rydberg excitation, interpreted as Rydberg-atom ionization creating charges that neutralize the device surface~\cite{supp}. The Rydberg spectral shift can be stabilized via both effects as close as \SI{90}{\um} to the device. 

\begin{figure}
\centering
\includegraphics[width=\columnwidth]{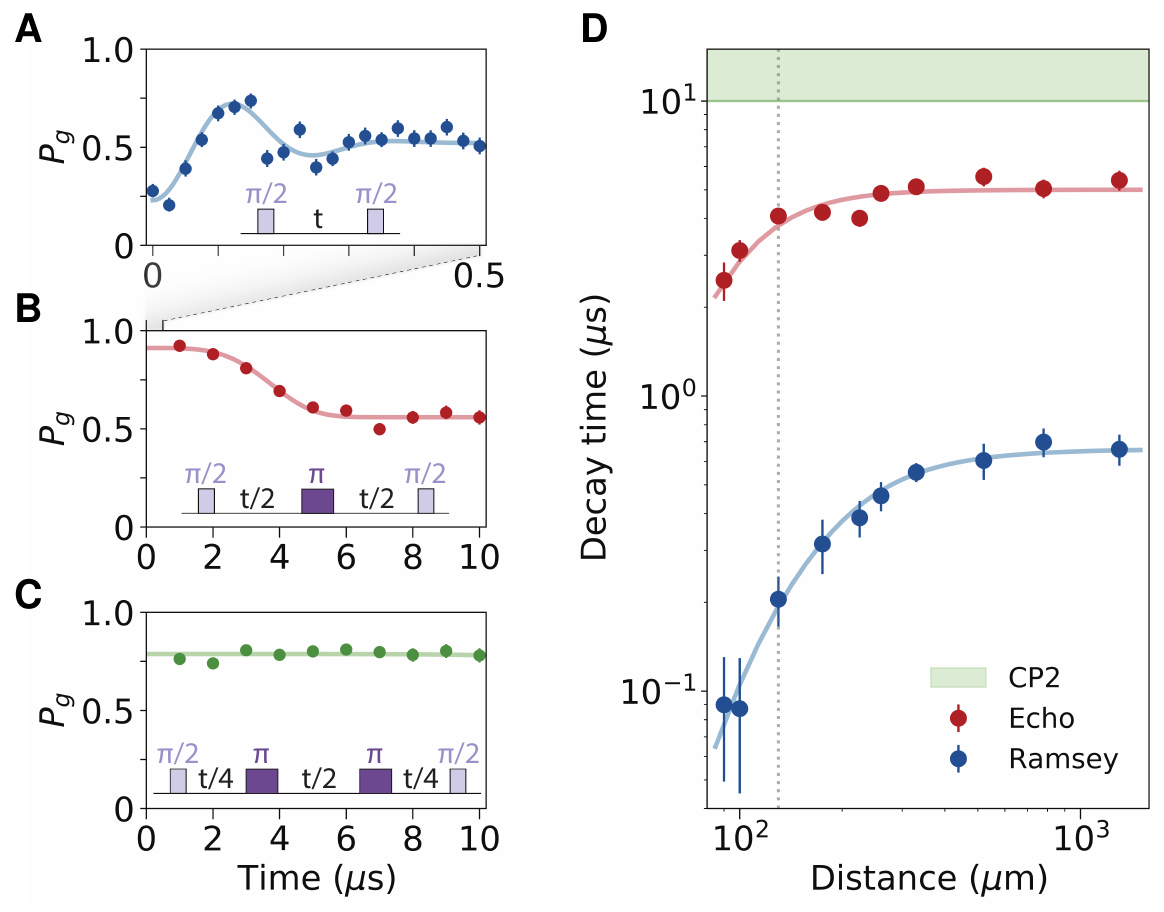}
\caption{\textbf{Dependence of single ground-Rydberg qubit coherence on distance from the device.} \textbf{(A)} $T_2^* = \SI{200(40)}{\nano\s}$ measured at \SI{130}{\um}. \textbf{(B)} $T_2 = \SI{4.1(2)}{\us}$ measured at \SI{130}{\um}. $T_2 \gg T_2^*$ implies that the electric field noise is quasi-static. \textbf{(C)} A Carr-Purcell $N=2$ decoupling sequence eliminates any decay within the measurement period, shown for \SI{130}{\um}. \textbf{(D)} $T_2$ (red) and $T_2^*$ (blue) as a function of distance from the device. $T_2^*$ is limited by electric field fluctuations, primarily from the background field. $T_2$ is limited by thermal sampling of the spatially-dependent electric field. The CP $N=2$ sequence eliminates decay at each measured distance and therefore extends the coherence to be $>\SI{10}{\us}$ (shaded green region). The data in (A-C) was taken at (dotted line).}
\label{fig:fig2}
\end{figure}

The temporal properties of the electric field from the device are probed by measuring the coherence of the ground-Rydberg qubit with various control sequences. By pulsing the \SI{420}{\nm} light, we apply a Ramsey sequence $\frac{\pi}{2}-t-\frac{\pi}{2}$ to extract $T_2^*$ as a function of distance from the device (Fig.~\refsub{fig:fig2}{A}). The Ramsey decay is found to be limited by detuning fluctuations of the Rydberg spectral shift caused by the electric field noise. With an echo pulse sequence (Fig.~\refsub{fig:fig2}{B}), the coherence time is extended by nearly an order of magnitude, implying that the electric field noise is quasi-static. While thermal sampling of the electric field gradient limits $T_2$, this source of decoherence can be further eliminated. Extending to a Carr-Purcell (CP) $N=2$ decoupling sequence (Fig.~\refsub{fig:fig2}{C}), no decay is observed within the \SI{10}{\us} evolution time at any measured distance. The scaling of $T_2$ and $T_2^*$ with distance is used to understand the sources of decoherence (Fig.~\refsub{fig:fig2}{D}). Fitting each to a model gives a background field standard deviation of \SI{0.012}{\V/\cm} and bounds the charge fluctuation of the device to $\le\SI{8}{e}$~\cite{supp}.

\begin{figure}[t]
\centering
\includegraphics[width=\columnwidth]{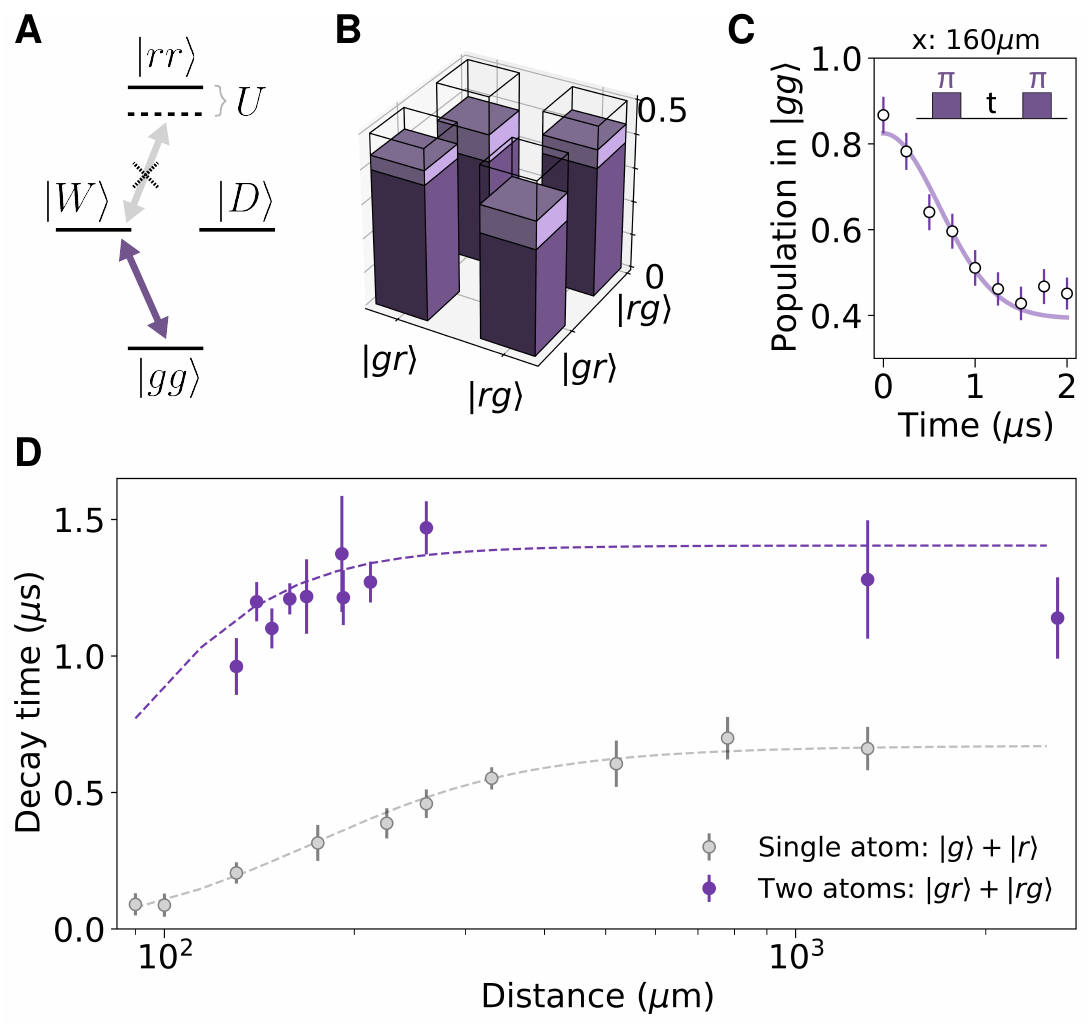}
\caption{\textbf{Entangled state preparation and lifetime.} \textbf{(A)} The Rydberg blockade interaction $U$ shifts the $\ket{rr}$ state and allows for preparation of the $\ket{W}=\frac{1}{\sqrt{2}}\left(\ket{gr}+\ket{rg}\right)$ state which can be rotated into $\ket{D} = \frac{1}{\sqrt{2}}\left(\ket{gr}-\ket{rg}\right)$ by applying a differential phase. \textbf{(B)} Extracted $\ket{W}$ state fidelity at \SI{170}{\um}, with population $\rho_{rg,rg}+\rho_{gr,gr}=0.87(5)$ and coherence $2|\rho_{gr,rg}|=0.78(7)$ with correction (light purple)~\cite{supp}. \textbf{(C)} $\ket{W}$ state lifetime at \SI{160}{\um} measured via a $\pi-t-\pi$ sequence. \textbf{(D)} The $\ket{W}$ state lifetime $T_W$, compared to single-atom $T_2^*$, is insensitive to common-mode electric field noise and is instead thermally limited.}
\label{fig:fig3}
\end{figure}

\begin{figure*}
\centering
\includegraphics[width=0.65\textwidth]{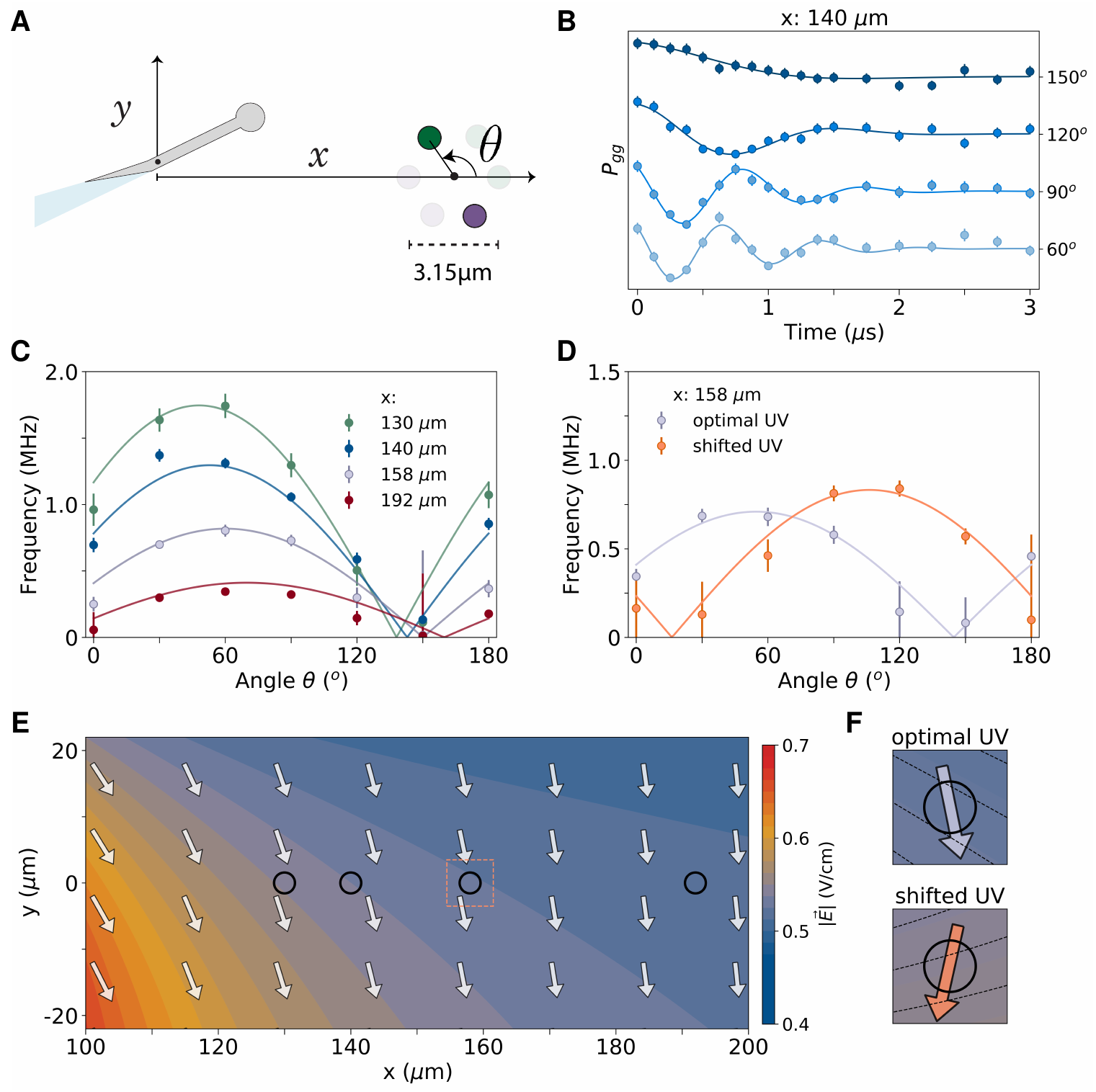}
\caption{\textbf{Entanglement-assisted sensing of the electric field environment.} \textbf{(A)} An atom-pair at varying $x$ distances from the device with angular orientation $\theta$ is used to sense the electric field gradient via oscillations between $\ket{W}$ and $\ket{D}$. \textbf{(B)} The $\ket{W}$ state lifetime measurement reveals the gradient-induced oscillations, shown at $x=$\SI{140}{\um} for different $\theta$ (offset for clarity). The entanglement assists in gradient sensing for distances where $T_W>T_2^*$. \textbf{(C)} Angular dependence of the oscillation frequency at different $x$. Simultaneous fitting of all curves gives a distance-independent background field in $x-y$ and a point-charge on the device. \textbf{(D)} A $\SI{200}{\nano\W}$ UV-power reduction from its optimized value causes a rotation (\SI{53}{\degree}) of the gradient, measured at $x=$\SI{158}{\um}. \textbf{(E)} Electric field contour plot showing $|\vec{E}|$ as reconstructed from (C) measurements (black circles) and direction (white arrows). The fiber-device junction is set to the origin. \textbf{(F)} Electric field contour plot reconstructed from (D) showing direction at the optimal UV value (gray arrow) and shifted UV value (orange arrow) using the same model from (C) to extract the changed parameters.}
\label{fig:fig4}
\end{figure*}

We next prepare and study entangled atom pairs at various distances from the device. Two atoms \SI{3.15}{\um} apart experience a Rydberg-Rydberg interaction $U$ in the blockade regime where $U\gg\Omega$, with $U \approx 2\pi\times\SI{685}{\MHz}$ and the single-atom Rabi frequency $\Omega \approx 2\pi\times\SI{3}{\MHz}$. The atoms are driven with a $\pi$-pulse from $\ket{gg}$ into the symmetric state $\ket{W}=\frac{1}{\sqrt{2}}\left(\ket{gr}+\ket{rg}\right)$ with a fidelity $\mathcal{F} = \frac{1}{2}\left( \rho_{rg,rg}+\rho_{gr,gr}+ 2|\rho_{gr,rg}| \right)$. Characterized following the method in \cite{levine18} and corrected for error and loss during preparation, the resulting fidelity  is \SI{0.83(4)}{} (uncorrected \SI{0.70(3)}{}) at \SI{170}{\um} (Fig.~\refsub{fig:fig3}{B}) and \SI{0.87(6)}{} (uncorrected \SI{0.73(3)}{}) at \SI{2.6}{\mm} from the device~\cite{supp}. This suggests that for the current system, the fidelity of the $\ket{W}$ state is not limited by proximity to the device. Furthermore, the lifetime of the $\ket{W}$ state $T_W$, measured through a pulse sequence $\pi-t-\pi$ between the $\ket{gg}$ and $\ket{W}$ (Fig.~\refsub{fig:fig3}{C}), 
is found to be $\sim$ 5 times longer than $T^*_2$ (Fig.~\refsub{fig:fig3}{D}), indicating that the entangled-state lifetime is not sensitive to the common-mode electric field fluctuations. $T_{W}$ remains fairly constant over the distances measured (Fig.~\refsub{fig:fig3}{D}) and is likely limited by thermal motion~\cite{supp}. 

Given its robustness, the $\ket{W}$ state can probe the local electric field gradient with greater sensitivity than single-atom measurements. The gradient creates an energy difference between $\ket{gr}$ and $\ket{rg}$ and thus a differential phase, leading to an oscillation between $\ket{W}$ and the anti-symmetric state $\ket{D}=\frac{1}{\sqrt{2}}\left(\ket{gr}-\ket{rg}\right)$. The oscillation frequency directly measures the gradient magnitude, observed via a $\pi-t-\pi$ pulse sequence. The direction of maximum gradient is detected by rotating the atom-pair by angle $\theta$ with respect to the $x$-axis (Fig.~\refsub{fig:fig4}{A}). Measured at four distances between $\SI{130}{\um}-\SI{192}{\um}$, using a UV power of \SI{400}{\nano\W} for each, the gradient maximum occurs at $\theta \sim\SI{55}{\degree}$ and its magnitude increases as the atoms approach the device (Fig.~\refsub{fig:fig4}{C}).
This differs from the expected \SI{0}{\degree} gradient orientation for a single point-charge displaced in $x$, but can instead be described by a minimal model that includes a distance-independent background electric field. A combined fit to the data in (Fig.~\refsub{fig:fig4}{C}) reveals a device charge $q = \SI{190(10)}{e}$ and background field $\mathbf{E} = (E_x,E_y) = (-0.02 (1), -0.51(3))\SI{}{\V/\cm}$~\cite{supp}. The oscillation frequency follows a point-charge gradient scaling of $r^{-5}$ along $\theta=\SI{0}{\degree}$ as well as $r^{-3}$ along $\theta=\SI{90}{\degree}$ as predicted by the model. 

Measuring the oscillation between $\ket{W}$ and $\ket{D}$ at various angles can be used to study the effect of UV light to understand the behavior illustrated in (Fig.~\refsub{fig:fig1}{C, inset}). At \SI{158}{\um}, shifting the UV power from the optimal value of \SI{400}{\nano\W} to \SI{200}{\nano\W} increases the spectral shift by \SI{6.7}{\MHz} and rotates the gradient by \SI{53(4)}{\degree} (Fig.~\refsub{fig:fig4}{D}). Assuming the UV power change modifies the fit parameters $q,E_x,E_y$ from Fig.~\refsub{fig:fig4}{C}, we find only a significant change in the background electric field: $\Delta E_x = \SI{-0.21(3)}{\V/\cm}$. A reconstruction of the changed electric field is illustrated in Fig.~\refsub{fig:fig4}{F}. However, far from the device at \SI{4}{\mm}, the electric field sensed by the atom does not vary with UV power. Measuring no change in the spectral shift for UV powers up to \SI{30}{\mW}, the electric field magnitude change can be bounded to $<\SI{0.05}{\V/\cm}$~\cite{supp}, much smaller than the observed $|\Delta E_x|$. This indicates that any UV light effect on the background electric field created by the surrounding glass cell is undetectable by the atom for this range of powers. Additionally, this step in UV power affects the charge on the device~\cite{supp}.

The influence of the UV illumination on the present surfaces can be understood by reconciling our observations. First, when choosing the optimal UV power at each distance, our data (Fig.~\refsub{fig:fig1}{C}) agrees with only a point-charge on the device. This is especially apparent for the measurements at negative $x$ distances where the atom is much closer to the fiber than the device and the spectral shift remains near zero~\cite{supp}. Second, a small deviation from the optimal UV power increases the total electric field $|E|$ at close distances but does not significantly change the charge $q$ on the device. Therefore, the fiber surface likely harbors a charge distribution that creates a background electric field when using a UV power that slightly deviates from the optimal. Remarkably, we can repeatably stabilize to a configuration where the fiber has no remaining charge by choosing the UV power that minimizes the spectral shift~\cite{supp}. This result enables the possibility of coherent control of Rydberg atoms near such devices and may also benefit other quantum platforms that are sensitive to surface charges~\cite{deleon21}.

These observations demonstrate the feasibility of interfacing micro- and nanoscale devices with Rydberg atom arrays. Potential improvements include using in-vacuum electrodes for better electric field control, minimizing the device size, or choosing a principal quantum number $n$ to balance the Rydberg interaction strength with electric field sensitivity~\cite{supp}. Importantly, even under present conditions, a Rydberg atom array placed $\sim\SI{250}{\um}$ away remains minimally perturbed by the device. At this distance, a ground-state atom can be entangled with a photon at the device~\cite{thompson13,tiecke14} and coherently transported to an atom array in the same tweezer~\cite{bluvstein22} as demonstrated from the nanophotonic cavity in \cite{dordevic21}. The transported atom can then be entangled with the Rydberg-atom quantum processor, opening a quantum optical channel via teleportation ~\cite{bennett93} to a photonic state or to a distant quantum processor. Such optical interfaces can also be utilized for fast, nondestructive readout for quantum error correction protocols~\cite{cong22,wu22}. More broadly, our work motivates integrating Rydberg atom arrays with other devices at this scale. For example, with properly designed decoupling sequences and charge stabilization, integration with a mesoscopic superconducting interface is possible, enabling the application of circuit QED techniques and exploration of novel hybrid systems~\cite{sorensen2004,pritchard2014,liu2022,petrosyan2009}.

\textbf{Acknowledgments:} We thank S. Ebadi, T.T. Wang, D. Bluvstein, T. Manovitz, G. Semeghini, and S. Evered for useful discussions, technical assistance, and sharing of laser light. We also acknowledge M. Greiner, R. Riedinger, and H. R. Sadeghpour for insightful discussions.
\textbf{Funding:}
This work was supported by
the Center for Ultracold Atoms, 
the National Science Foundation,
AFOSR MURI, 
DOE QSA Center
and ARL CDQI. 
B.G. acknowledges support from the DOD NDSEG.
The device was fabricated at the Harvard CNS (NSF ECCS-1541959).
\textbf{Author contributions:} 
P.L.O., I.D. and E.G-S. performed the measurements. P.L.O., I.D., B.G., and E.G-S. analyzed data. B.G. and T.\DJ. developed the analytical coherence models. P.S. and P.L.O. fabricated the device.
All work was supervised by V.V. and M.D.L. All authors 
discussed the results and contributed to the manuscript. 
\textbf{Competing interests:} 
V.V. and M.D.L. are co-founders and shareholders of QuEra Computing.
\textbf{Data and materials availability:} All data needed to evaluate the conclusions in the
paper are present in the paper and the supplementary materials.



\clearpage

\renewcommand{\thefigure}{S\arabic{figure}}
\renewcommand{\thetable}{S\arabic{table}}
\setcounter{equation}{0}
\setcounter{figure}{0}
\setcounter{table}{0}

\begin{center}
{\LARGE Supplementary Materials}
\end{center}
{\let\newpage\relax\maketitle}

\tableofcontents

\section{Setup details}
\label{sec:setup details}

The setup consists of a vacuum chamber with a glass cell on a translation stage [LinTech LTE 170SERIES-4] which can be moved in and out along the $x-$axis relative to the optical beams and magnetic field coils. This allows the distance between the atoms trapped in the tweezers and the SiN device, attached to the vacuum chamber, to be varied with precision of about $1-2  \SI{}{\mu m}$ for distances within the field of view of the imaging system ($ < \SI{225}{\mu m}$). For distances outside of the field of view of the imaging system, the movement is calibrated given the pitch of the translation stage main screw with precision of $\sim \SI{15}{\mu m}$. This calibration is corroborated within our field of view. Two objectives [Mitutoyo G Plan Apo 50X] on either side of the glass cell focus the tweezer, optical pumping, and Rydberg beams. The tweezers use \SI{815}{\nm} light. The atoms in the tweezers are loaded from a MOT, followed by polarization-gradient cooling, after which the atoms are pumped into the $|F, m_F\rangle = |2,2\rangle$ Zeeman level of the ground $\ket{5S_{1/2}}$ state via optical pumping. The optical pumping uses a focused circularly-polarized \SI{780}{\nm} beam addressing the $F=2$ manifold, while the $F=1$ manifold is addressed by the MOT repumper beam. The tweezers are turned off during Rydberg excitation.

Ground-state atoms are excited to the $\ket{70S, {J=1/2}, {m_J = 1/2}}$ state by a two-photon transition using \SI{420}{\nm} and \SI{1013}{\nm} light detuned by \SI{2}{\GHz} from the intermediate $|F,m_F\rangle = |3,3\rangle$ level in the $6P_{3/2}$ manifold. The \SI{420}{\nm} light comes from an injection-locked TopGan \SI{418}{\nm} diode seeded by frequency-doubled Ti:Sapphire laser [M Squared, \SI{15}{\W} pump]. The frequency of the Ti:Sapphire laser is stabilized by locking the fundamental mode to an ultra-low expansion (ULE) reference cavity (notched cylinder design from Stable Laser Systems), with finesse $\mathcal{F} = 30000$ at \SI{840}{\nm}. The \SI{1013}{\nm} light comes from an external-cavity diode laser [Toptica DL Pro], which is locked to the same reference cavity ($\mathcal{F} = 50000$ at \SI{1013}{\nm}). It is used as a seed for an injection-locked Toptica AR \SI{1060}{\nm} diode, which then seeds a MogLabs \SI{1013}{\nm} TA.

For the single-atom data, these beams are focused from the front and the back objectives, such that they counter-propagate to minimize Doppler shifts. The single-photon Rabi frequencies in this case are $\Omega_{420} = 2\pi\times\SI{170(5)}{\MHz}$ and $\Omega_{1013} = 2\pi\times\SI{73(3)}{\MHz}$ with spot sizes of $w_{420} \approx  \SI{3}{\mu \m}$ and $w_{1013} \approx  \SI{1.5}{\mu \m}$ $1/e^2$ radius. For the two-atom data, the Rydberg excitation beams were expanded with spot sizes of $w_{420} \approx  \SI{11.5}{\mu m}$ and $w_{1013} \approx  \SI{16}{\mu \m}$ $1/e^2$ radius, while the single-photon Rabi frequencies are kept roughly unchanged. To enlarge the Rydberg excitation beams it became necessary to make them co-propagating, focused via a doublet lens which replaced one objective, such that now both counter-propagating to the optical tweezers.

\begin{figure}[h!]
\center
\includegraphics[width=0.9\columnwidth]{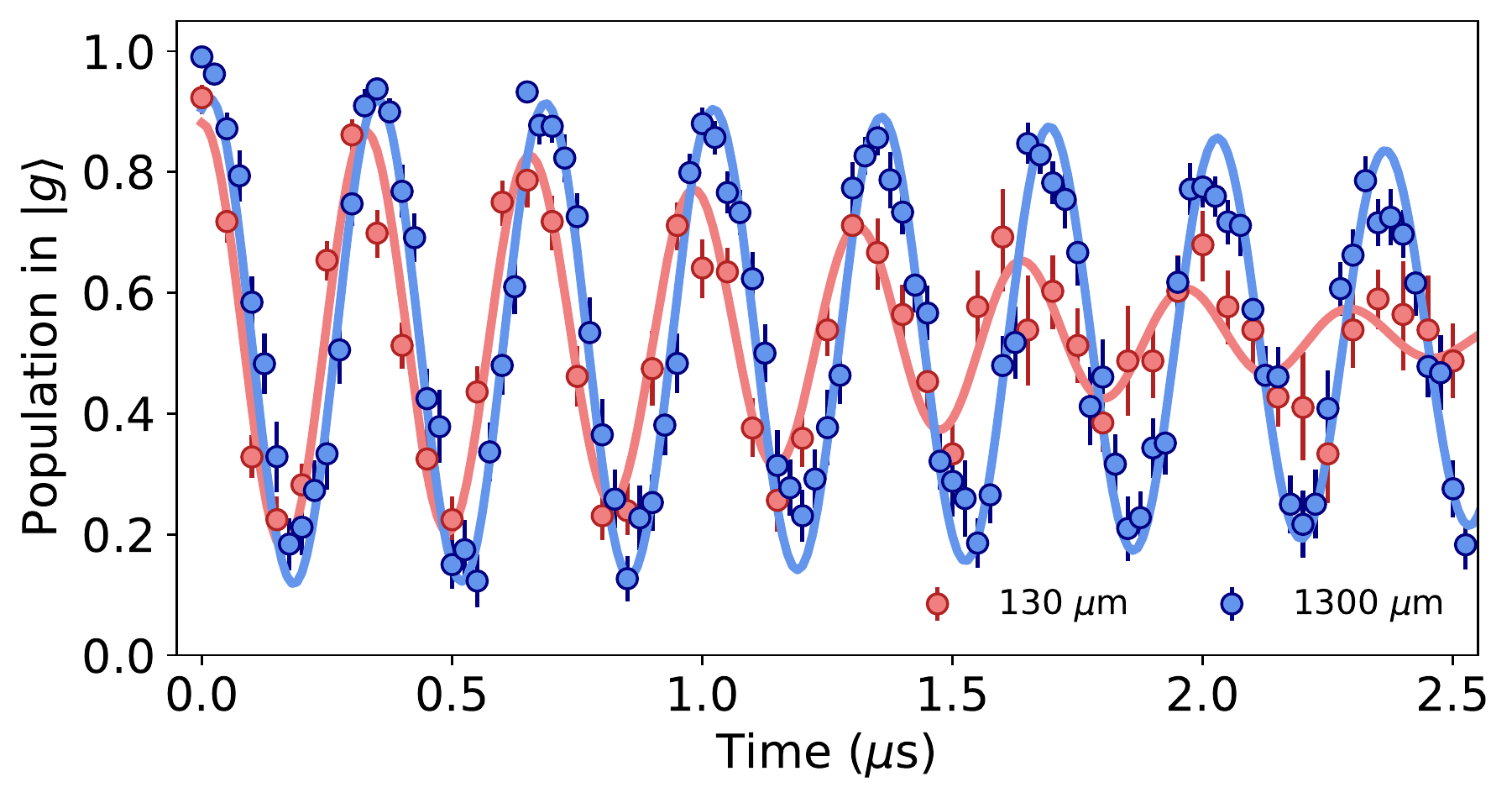}
\caption{Single-atom Rabi oscillations near (\SI{130}{\mu \m}) and far (\SI{1300}{\mu \m}) from the device. The decay time $\tau$ decreases close to the device with $\tau_{130} = \SI{1.6(1)}{\mu s}$ and $\tau_{1300} = \SI{4.9(2)}{\mu s}$.}
\label{fig:single_atom_rabi}
\end{figure}

The single-atom Rabi oscillations are plotted in Fig.~\ref{fig:single_atom_rabi} for two distances ($130$ and $\SI{1300}{\mu m}$). The solid lines are fits of the form: 
\begin{equation}
f(t) = a \cos(2\pi f t ) e^{-\left( t/\tau \right)^2}+c    
\end{equation}
where we find only the decay time $\tau$ is significantly different and both curves agreeing with $\Omega=2\pi\times\SI{3}{\MHz}$ within errorbars. Assuming that at far distances detuning noise from electric field fluctuations is small, the decay of the Rabi oscillations corresponds to \SI{1.5}{\%} Rabi frequency fluctuations. This could be due to sampling of different Rydberg laser intensities due to beam pointing given the small beam sizes. At closer distances, the Rabi decay time is limited by detuning noise. In particular, for the $\SI{130}{\mu\m}$ distance, quasi-static detuning noise with \SI{2}{\MHz} standard deviation could explain the observed decay time.

\section{Stabilizing the Rydberg spectral shift}

\subsection{UV power tuning}
\label{subsection:effect_of_uv_1}

\begin{figure}
\center
\begin{overpic}[width=0.75\columnwidth]{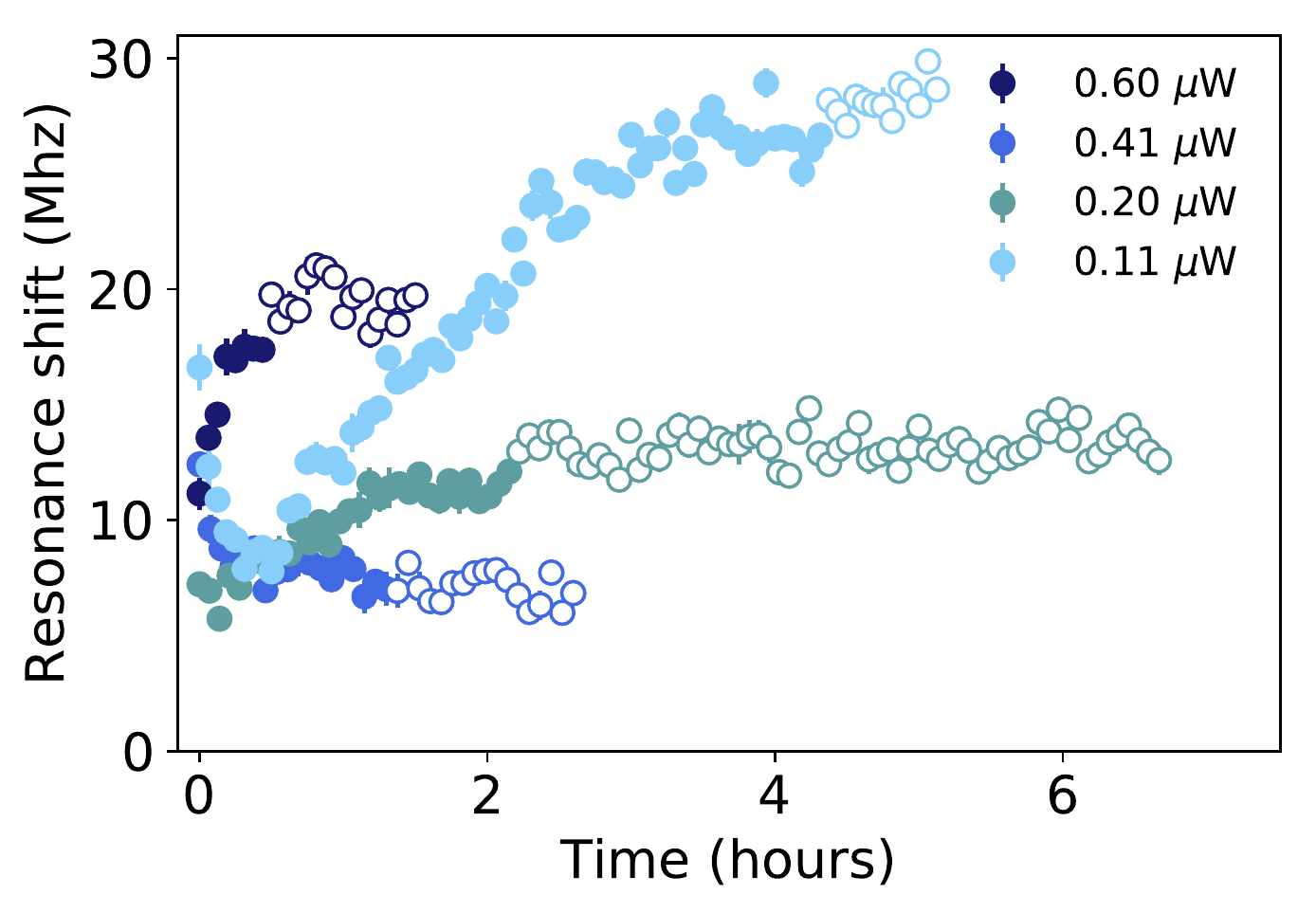}
\put(-1,65){\textbf{(a)}}\end{overpic}
\begin{overpic}[width=0.75\columnwidth]{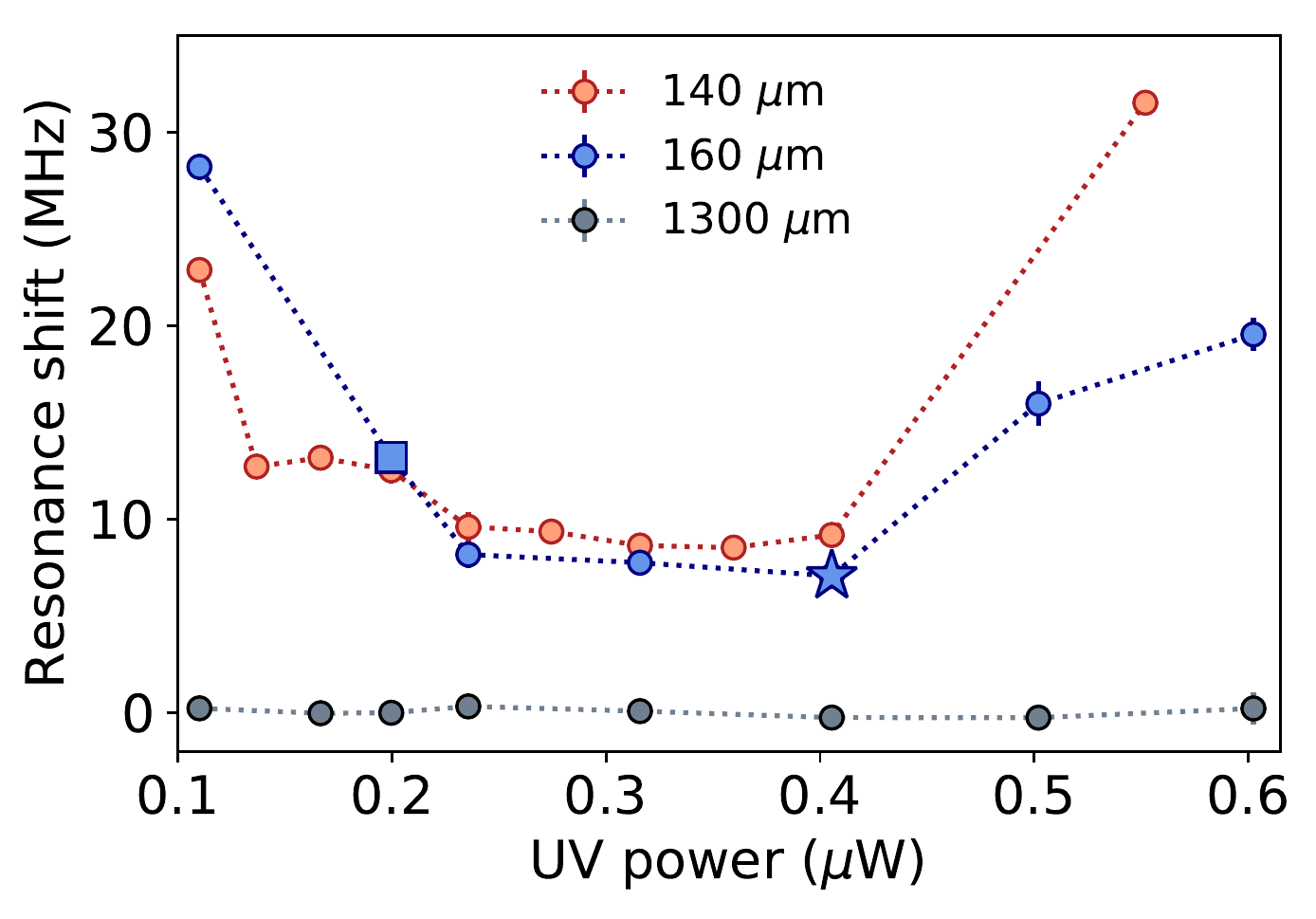}
\put(-1,65){\textbf{(b)}}\end{overpic}
\caption{UV power tuning. \textbf{(a)} Rydberg spectral shift settling after changing the UV power at 160 $\mu$m. \textbf{(b)} Settled shift at different distances as a function of the UV power. The 160 $\mu$m curve is generated from the averages of white-faced points in (a). For Fig. 4(D) in the main text the UV power is changed from the optimal value (star) to the shifted value (square).}
\label{fig:shift_vs_uv}
\end{figure}

We observe that in order to stabilize the Rydberg spectral shift close to the device, a small amount of UV power is required. We shine a \SI{395}{\nm} diode [Thorlabs M395L5] on the entire system from the side as depicted in Fig. 1 of the main text. After changing the power of the UV diode, the spectral shift takes about an hour to settle. This is shown in Fig.~\ref{fig:shift_vs_uv}(a) for different UV powers at 160 $\mu$m. The settled Rydberg spectral shifts are denoted with white-face symbols and the averages of those points are plotted as a function of the power in Fig.~\ref{fig:shift_vs_uv}(b). Surprisingly, we find that there is a minimum in the Rydberg spectral shift as a function of UV power observed at distances closer than $\sim\SI{800}{\mu m}$ from the device. Our observations in Section~\ref{sec:effect_of_uv_on_gradient} further imply that tuning the UV power primarily affects the local background electric field at close distances. For larger distances, we see no effect of the Rydberg spectral shift when the UV power is changed in this range. As noted in the main text, we interpret this to signify the presence of another surface close to the device (e.g. the tapered fiber) which is affected differently by the UV power. The minimum is then due to the sum of the two electric fields for surfaces with a different UV power dependence.

\begin{figure}
\center
\begin{overpic}[width = 0.75\columnwidth]{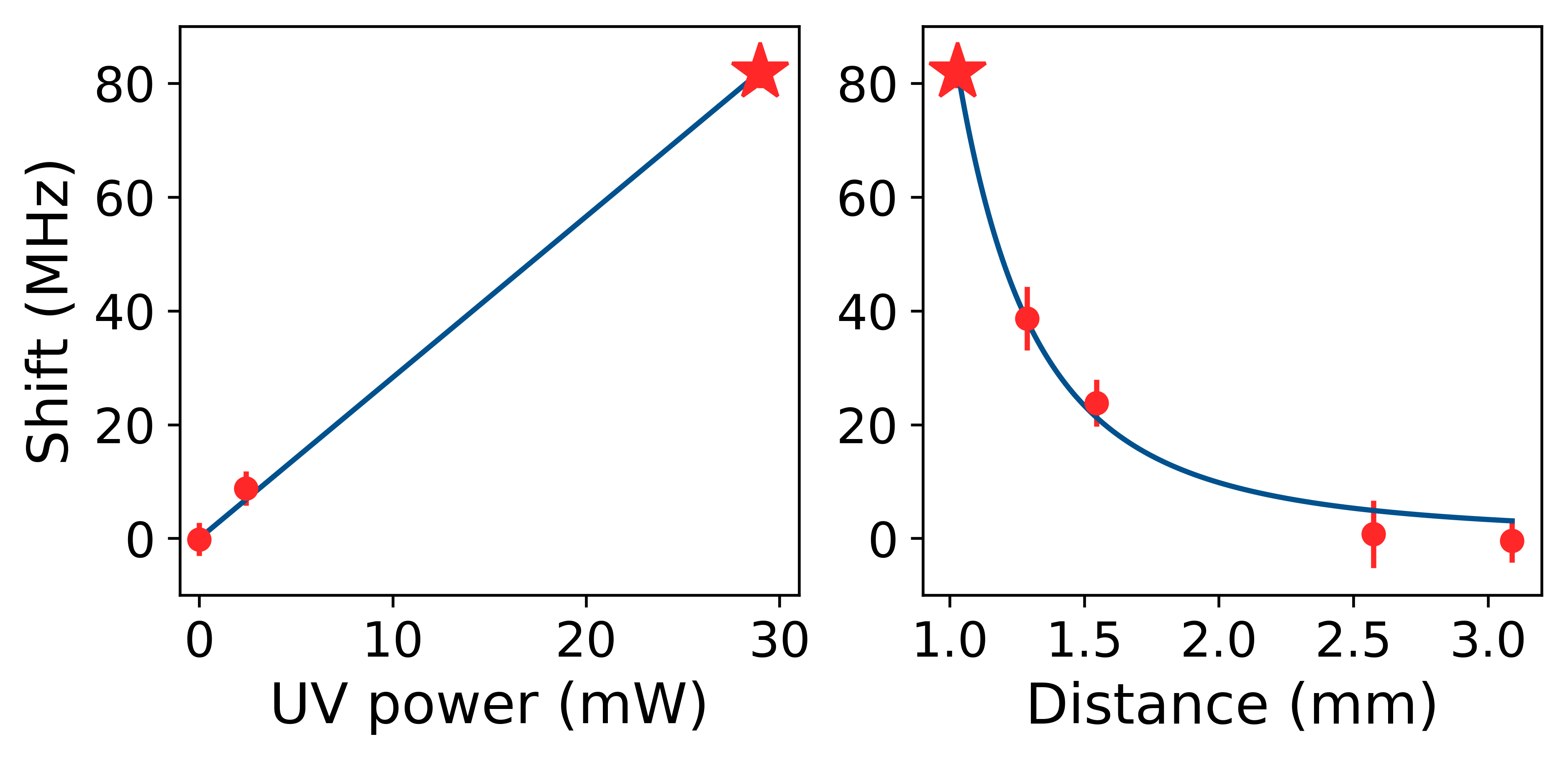}
\put(4.4,48){\textbf{(a)}}\put(52,48){\textbf{(b)}}
\end{overpic}
\caption{Effect of large UV power. \textbf{(a)} Measured spectral shift as a function of UV power at a distance of $\SI{1}{\mm}$ from the device. A single parameter linear fit is included as a guide to the eye. \textbf{(b)} Spectral shift as a function of distance from the device at a UV power of $\SI{29}{\mW}$. The line is a fit to a point-charge model with $Q=3.4(3)\times10^{4}$\SI{}{e} charges and $E_x=\SI{0.08(4)}{\V/\cm}$. The points marked by in a star are the same data point.}
\label{fig:largeUV}
\end{figure}

Larger UV powers further increase the Rydberg spectral shift. This is illustrated for a distance of 1mm in Fig.~\ref{fig:largeUV}(a). Keeping the power at 29 mW and varying the distance from the device, we plot the Rydberg spectral shift in (Fig.~\ref{fig:largeUV}(b)). Using a point-charge model, we fit $\approx 34,500$ elementary charges at the device and a background field of $E_x=0.08(4)$ V/cm. This charge is more than two orders of magnitude larger than the $126(11)$ charges we obtain when using low UV powers that minimize the spectral shift, as seen in the main text. Further, at a distance of \SI{3.9}{\mm}, UV powers in the range $0$-$\SI{30}{\mW}$ do not significantly change the spectral shift. This bounds the maximum change with UV power of the background electric field from the glass cell that the atom can sense to be $|\Delta \mathbf{E_0}|_{\rm{max}}\sim 0.05$ V/cm. 

\subsection{Effect of Rydberg production}
\begin{figure}
\center
\begin{overpic}[width=0.75\columnwidth]{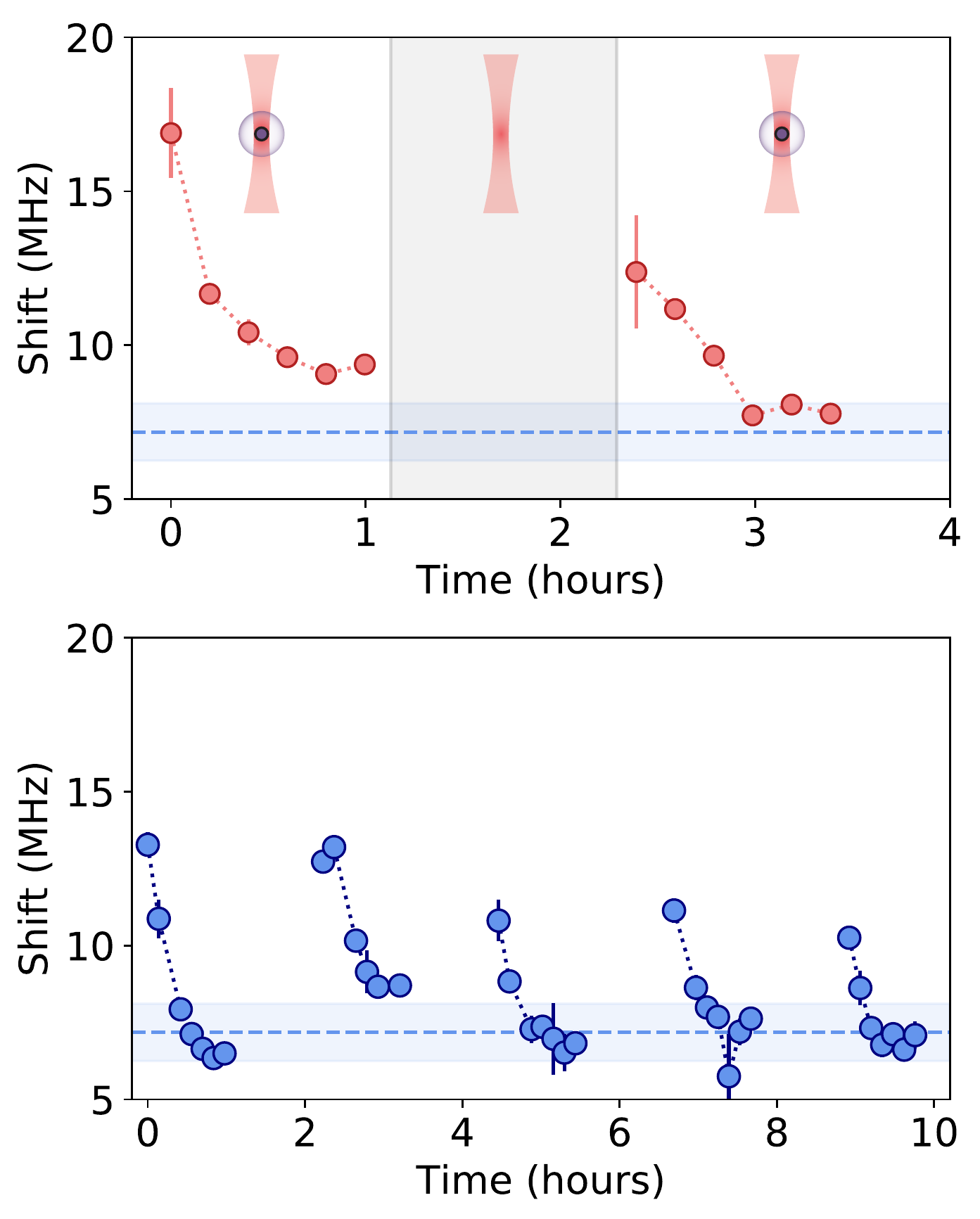}
\put(0,100){\textbf{(a)}}\put(0,50){\textbf{(b)}}\end{overpic}
\caption{Rydberg production studied at $\SI{130}{\mu m}$. \textbf{(a)} After the initial settling of the Rydberg spectral shift over the first hour, the experiment is run without an atom in the tweezer. When the single atom is again loaded, the spectral shift settles. The dashed blue line and shaded region is the average settled value and standard deviation from (b). \textbf{(b)} The Rydberg spectral shift monitored between periods of running the experiment with no Rydberg excitation by not sending the 420nm or 1013nm light. The dashed blue line and shaded region is the average settled value and standard deviation respectively. }
\label{fig:ryd_prod}
\end{figure}

In addition to UV illumination, we observe that a constant rate of exciting Rydberg atoms close to the device is required to stabilize the spectral shift. When Rydberg atoms are not being excited, the Rydberg spectral shift increases to higher energies signifying that the electric field has increased. When Rydberg atoms are excited again, the shift settles back to its original value, typically in less than an hour. This effect is illustrated in Fig.~\ref{fig:ryd_prod}. In Fig.~\ref{fig:ryd_prod}(a), we demonstrate the spectral shift sensitivity to single atoms. After the spectral shift initially settled for this distance of $130\mu$m, the experimental sequence was run without an atom in the tweezer for a period of time. Afterwards, we returned to exciting single atoms to the Rydberg state and observe that the spectral shift returns to its original settled value. We interpret this behavior as the result of ionization by the \SI{815}{\nm} tweezer light creating free charges that partially neutralize the charge on the device. Therefore, this experimental demonstration suggests that the settling behavior cannot be an effect of any background atoms. Instead, the charge on the device seems sensitive to the charges from single ionized atoms. Note that similar effects were observed in Rydberg EIT experiments near a quartz surface \cite{shaffer16}. 

From this measurement we can approximate the rate at which the free charges reach the device to lower the overall charge. Using a simple linear fit to the slope of the second half of Fig.~\ref{fig:ryd_prod}(a) we estimate a rate of $1.7(1)$ charges per minute ($\sim\SI{0.03}{Hz}$) which we can compare to our experimental duty cycle of $\sim \SI{1.3}{Hz}$. In Fig.~\ref{fig:ryd_prod}(b), we study the effect of Rydberg production on the spectral shift over many hours between periods of not sending the Rydberg excitation beams and observe the same characteristic behavior. While Fig.~\ref{fig:ryd_prod}(a-b) were both taken for a distance of 130$\mu$m from the device, we observe similar behavior for distances less than $\sim 500\mu$m.

\begin{figure}
\centering
\begin{overpic}[width=0.75\columnwidth]{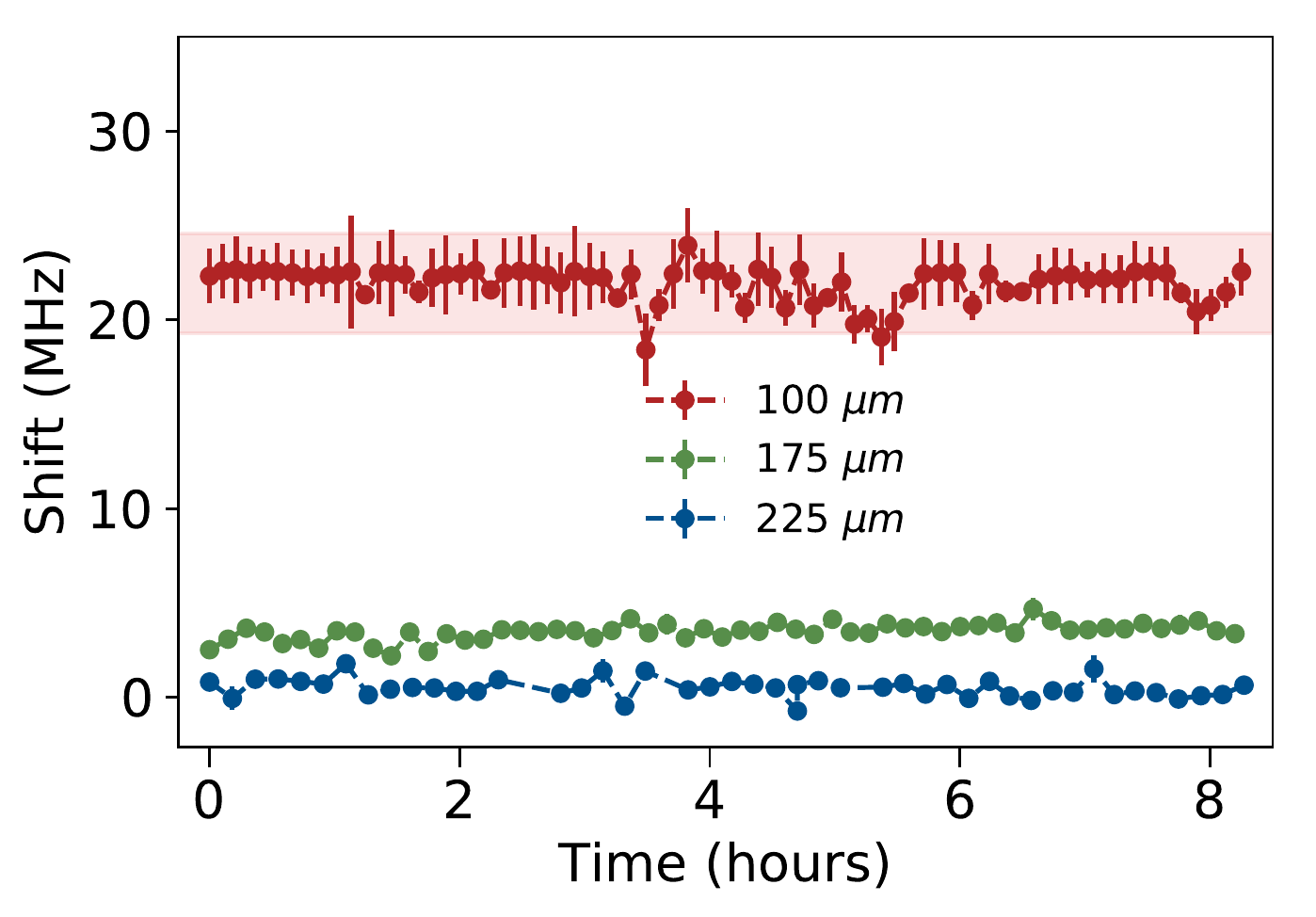}
\put(0,65){\bf (a)}\end{overpic}
\begin{overpic}[width=0.75\columnwidth]{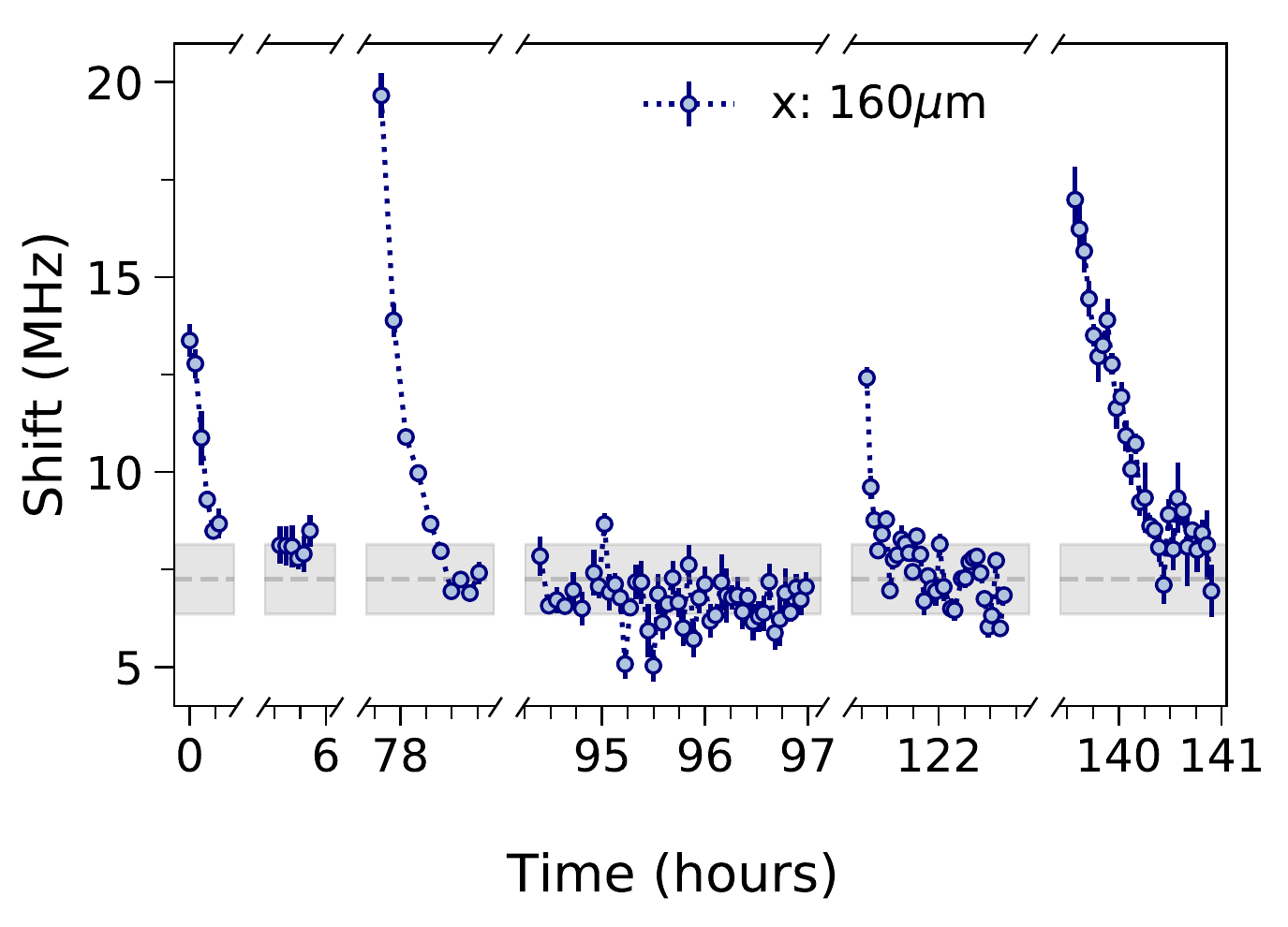}
\put(0,70){\bf (b)}\end{overpic}
\caption{Rydberg spectral shift stability in time. \textbf{(a)} Stability shown over 8 hours at several distances after the initial settling of the spectral shift. The shaded area here is the Rabi-broadened linewidth of 3MHz. \textbf{(b)} Reproducibility over 6 days at $\SI{160}{\mu m}$ where many settling periods can be seen, each after a period of either not running the experiment or not exciting Rydberg atoms. The shaded area is one standard deviation ($\pm$ 0.9 MHz) around the average settled value ($7.2$ MHz) of all curves.} 
\label{fig:line_stability}
\end{figure}

\subsection{Rydberg spectral shift stability and reproducibility}
After the Rydberg spectral shift has been minimized and settled, it remains stable. This is shown in Fig.~\ref{fig:line_stability}(a), where the Rydberg spectral shift is monitored over several hours at different distances. Although the magnitude of fluctuations of the spectral shift increases as we approach the device, it remains below the Rabi-broadened linewidth of 3 MHz as close as 90 $\mu$m. 

We observe that the spectral shift and thus the electric field environment is reproducible. For example, we have monitored the Rydberg spectral shift over the course of a week during normal operation as shown in Fig.~\ref{fig:line_stability}(b). Several settling behaviors are observed throughout the week after periods of no Rydberg production. We observe that the Rydberg spectral shift repeatedly returns to within a linewidth of the average settled value.

Remarkably, we observe similar behavior even over longer periods of time. For example, the data in Fig. 1 and Fig. 4 in the main text were taken about 6 months apart. We plot the Rydberg spectral shifts at the distances used in Fig.~\ref{fig:fig4_line}. We see that the Rydberg shifts lie within the Rabi broadened linewidth of 3 MHz from each other at each given distance. Furthermore, each Fig. 4 data point lies on the point-charge fit from Fig. 1, except for the 130 $\mu$m data point. This data point in fact was taken at a UV value that slightly deviated from its optimum value, as explained below. The ability to bring the Rydberg spectral shift back to this curve even after 6 months indicates that by tuning the UV power and producing Rydbergs at a constant rate, we can stabilize to the same charge environment repeatedly.

\section{Electric field from the device}

\subsection{Point-charge model}
We consider a minimal model in which the SiN device acts as a point charge $q$ and there is a homogeneous background electric field $\mathbf{E_0}$. The point-charge approximation is appropriate given that the device is $\SI{31.5}{\mu m}$ long and we explore distances from $\SI{90}{\mu m}$ to \SI{3}{\mm}. Our model assumes that the charge is located in the middle of the device Fig.~\ref{fig:linecharge}(a). The energy shift of the Rydberg state is:
\begin{equation}
    \nu(r) = \frac{1}{2} \alpha \left|\frac{kq}{r^2}\mathbf{\hat{r}} + \mathbf{E_0}\right|^2
    \label{eq:line_shift}
\end{equation}
where $\alpha$ is the polarizability of the Rydberg state, which for $\ket{70S}$ is $\alpha \approx \SI{534}{\MHz/(\V/\cm)^2}$, and $\mathbf{r}$ is the distance between the Rydberg atom and the device. The spectrum we measure in Fig. 1 in the main text is referenced to our farthest distance. As a result we measure a differential shift given by:
\begin{equation}
\Delta \nu \equiv \nu(r)-\nu(\infty) = \frac{1}{2} \alpha  \left( \frac{(kq)^2}{r^4}+ \frac{2kq}{r^2}\mathbf{E_0}\cdot\mathbf{\hat{r}}  \right)
\label{eq:differential_line_shift}
\end{equation}
This model fits well to the measured spectral shift, which exhibits a clear $\propto r^{-4}$ scaling with distance. Assuming a 1D geometry, we extract values of $q = 126(11)$ charges and a small $E_x = \SI{0.05(1)}{\V/\cm}$ field. This value of $q$ is extracted by assuming the charge is located at the center of the device, however we fit for values $80-180$ charges if assumed at either end of the device. This point-charge model is used throughout the main text for single atoms and is expanded to a 2D geometry when using two-atoms.

\subsection{Line charge model}
\begin{figure}
\center
\begin{overpic}[width=0.75\columnwidth]{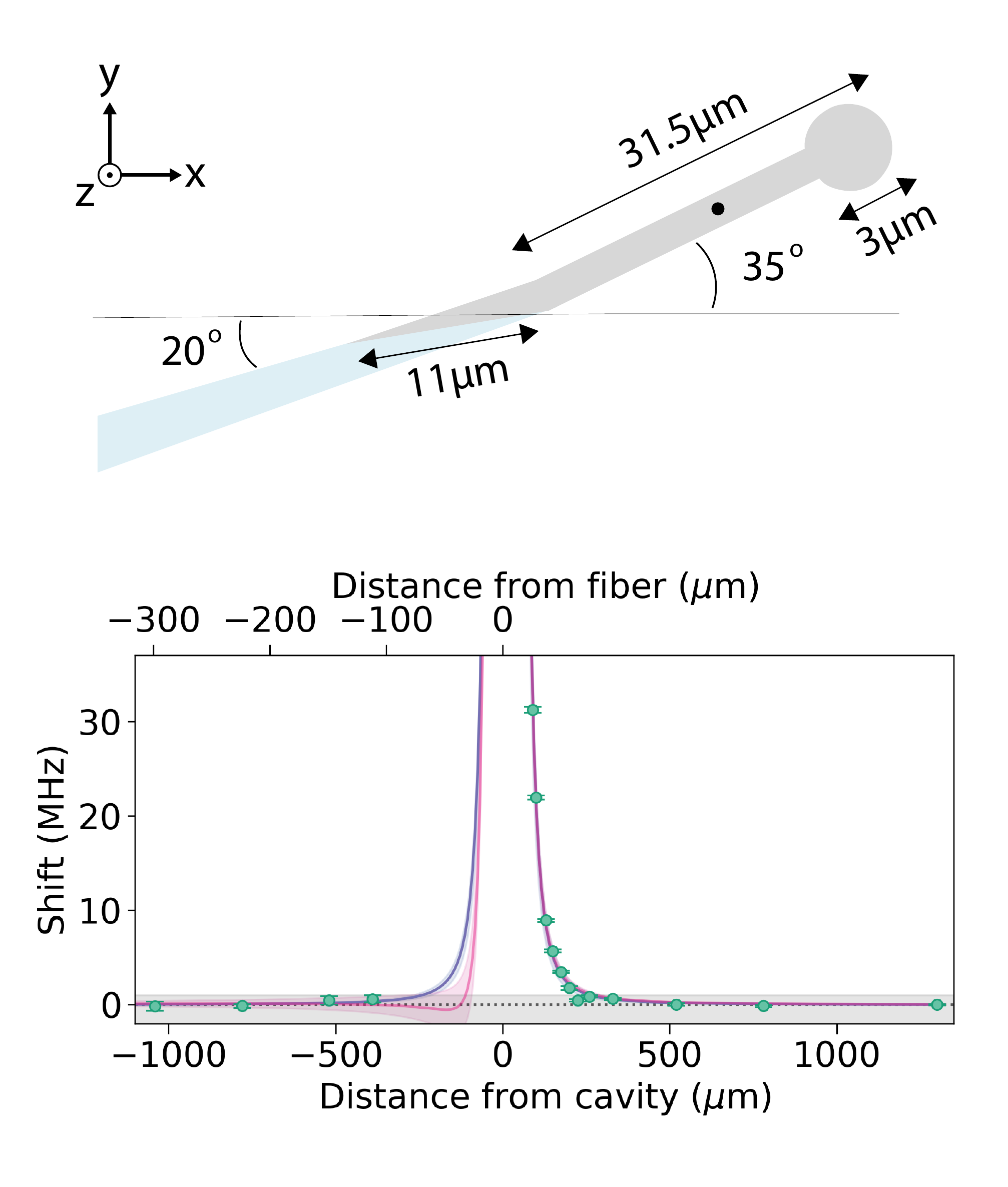}
\put(0,100){\textbf{(a)}}\put(0,50){\textbf{(b)}}\end{overpic}
\caption{SiN device attached to a silica fiber. \textbf{(a)} Diagram showing the dimensions and angles of the device with respect to the $x-$axis. \textbf{(b)} Measured spectral shift (green circles) as a function of distance from both the device and the fiber surface. We fit the measured shifts to a point-charge (blue) and a line-charge (pink) model, finding good agreement with charges on the SiN device and not on the fiber.}
\label{fig:linecharge}
\end{figure}

We can evaluate the point-charge approximation by going one step further and constructing a model which takes into account the silica tapered fiber that the SiN device is attached to, as shown in Fig.~\ref{fig:linecharge}(a). We assume a semi-infinite linear charge density $\lambda_{\rm fib}$ for the fiber with its end-point at the origin and at -\SI{20}{\degree} with respect to the $x-$axis and a second finite linear charge density $\lambda_{\rm dev}$ for the device starting at the origin, with length $L=\SI{31.5}{\micro\meter}$ and at \SI{35}{\degree} with respect to the $x-$axis. This model allows us to differentiate between the possible surface charge characteristics of each material. Here we assume a 2D geometry with a fixed perpendicular field along $\hat{y}$ to $E_y=\SI{-0.51}{\volt / \cm}$ (in accordance with our gradient measurements in Section \ref{sec:gradient}). The fit is shown in Fig.~\ref{fig:linecharge}(b) and returns the parameters: $\lambda_{\rm dev} = \SI{4.8(2)}{e/\micro\meter}$, $\lambda_{\rm fib} = \SI{-0.01(1)}{e/\micro\meter}$, and $E_{0,x}=\SI{0.06(1)}{\volt/\cm}$. These parameters agree with a total number of $Q=\SI{152(5)}{e}$ charges equally distributed over the length of the device and no charges on the fiber. 

For the linear charge distribution model Fig.~\ref{fig:linecharge}(a), the fit takes into account all points from Fig. 1 in the main text. The points with negative values correspond to points where the distance between the atom and the device is larger than the distance between the atom and the fiber surface given the geometry of the setup. Our observation of no spectral shift up to $\SI{100}{\micro\meter}$ from the fiber surface reinforces the conclusion that the vast majority of the charges are confined to the SiN device. This allows us to simplify the charge distribution as a point-charge in the main text. Further investigation is required to explain how working at the optimal UV power zeros the tapered fiber of charge, as it could help the design of future photonic link architectures for Rydberg atom arrays.

\section{Single-atom coherence}
The presence of an electric field which has both a spatial dependence and time fluctuations will cause decoherence between the Rydberg state and the ground state, since only the Rydberg state is sensitive to the electric field. The time fluctuations will cause detuning noise on the ground-Rydberg transition. A thermal atom will sample different electric fields, also causing detuning noise.

\subsection{Ramsey and Echo decay measurements}
\label{sec:t2_to_sigma_delta}
To study the coherence of a Rydberg atom in this environment, we use a Ramsey sequence $\pi/2 - t - \pi/2$ and measure the population in the ground state as a function of time $P_g(t)$. We pulse on the 420 nm beam during the $\pi/2$ pulses while the 1013 nm beam is kept on throughout the experiment, similar to \cite{levine18}. Since it is far-detuned, the 1013 nm beam does not cause scattering. For a general two-level system, the Ramsey measurement results in a measurement of the differential phase accumulation $\Delta \phi$ between $\ket{g}$ and $\ket{r}$. The evolution of the ground state population as a function of the hold time between the two $\pi/2$ pulses is then: 
\begin{equation}
    P_g(t) = \frac{1}{2} \left(1 + \cos( \Delta \phi(t)) \right)
\end{equation}
In our model (see Section~\ref{sec:single-atom-ramsey}), the main decoherence effect is detuning noise from electric field fluctuation, such that $\Delta \phi = \delta t$ for detuning $\delta$. We can assume $\delta$ is normally-distributed around $\delta_0$, which in our case is the light shift of the 420 nm beam, with a standard deviation $\sigma_{\delta}$ and that the noise is quasi-static: at each realization of the experiment we probe one value from this distribution. As we average over the distribution, we obtain a decay:
\begin{align}
    P_g(t) = & \int \frac{1}{2} \left(1 + \cos(\delta t) \right) e^{-(\delta-\delta_0)^2/2\sigma^2_{\delta}} d\delta \nonumber \\
    = & \frac{1}{2} \left( 1+\cos(\delta_0 t) e^{-t^2 \sigma^2_{\delta}/2} \right)
\end{align}
We see that the characteristic decay time is:
\begin{equation}
    \tau = \frac{\sqrt{2}}{\sigma_{\delta}}
\end{equation}
Therefore, to obtain the decay time $\tau$ we use a fitting function of the form:
\begin{equation}
    f(t) = a \cos(2 \pi f t + \phi_0) e^{-\left( t/\tau \right)^2}
\end{equation}

The echo pulse sequence is $\pi/2 - t - \pi - t - \pi/2$. Again, the population in the ground state is measured and the decay time $\tau$ is obtained by a fit of the form: 
\begin{equation}
    f(t) = a \cos(2 \pi f t + \phi_0) e^{-\left( t/\tau \right)^4}
\end{equation}
Since the quadratic decay is cancelled out by the echo sequence, the echo signal has the next order in the time dependence.

\subsection{Theoretical model}
Given a point charge distribution and a background electric field, as in Eq.~\eqref{eq:line_shift}, we can develop a model for the dependence of $T_2^*$ and $T_2$ on the distance from the device. For simplicity we will work with angular frequencies, $\omega$, such that the energy shift from the electric field is:

\begin{equation}
\delta \omega(r) = 2\pi\delta \nu(r) \approx \pi \alpha |\frac{kq}{r^2}\mathbf{\hat{r}}+\mathbf{E_0}|^2  
\label{eq:delta_omega}
\end{equation}

At each distance $r$, the spatial dependence of the energy shift is approximately linear with distance around the position of the atom. We use the local energy gradient $\nabla\delta \omega(r)$ to simplify our model of dephasing. Additionally we assume that the motional state of the atom is described by a thermal Maxwell-Boltzmann distribution of the initial position $\mathbf{r_0}$ and initial velocity $\mathbf{v_0}$. We find that this results in the Rydberg state experiencing a time-dependent detuning given by:

\begin{equation}
    \delta \omega(t) = \delta \omega_0 + \nabla \delta \omega \cdot (\mathbf{r}_0 + \mathbf{v}_0t + \frac{\mathbf{a}}{2}t^2)
    \label{eq:timedepdet}
\end{equation}
where $\delta \omega_0$ is the initial detuning and the acceleration $\mathbf{a}$ is the result of the force from the electric field gradient and is given by:  
\begin{equation}
    \mathbf{a} =  \frac{\hbar}{m}\nabla \delta \omega
    \label{eq:acc} 
\end{equation} 

It is only when the atom is in the Rydberg state that it accelerates and in each pulse sequence the atom will exist in a superposition of the Rydberg and ground state for some time. In order to account for the state-dependent force, we can define a function $f_{+,-}(t)$ such that:

\begin{equation}
f_{+}(t) =  1-f_{-}(t)
\end{equation}
which can be written into the time-dependent detuning of the Rydberg state:
\begin{equation}
    \delta \omega_{+, -}(t) = \delta \omega_0 + \nabla \delta \omega \cdot (\mathbf{r}_0 + \mathbf{v}_0t +\int^t_0\int^{t}_0 \mathbf{a}f_{+,-}(t')dt' dt'') 
\end{equation}
We can then calculate the relative phase acquired by the Rydberg state and ground states:
\begin{equation}
    \phi(t) = 2\pi\int_{0}^\tau f_{+}(t)\delta \omega_{+}(t)  - f_{-}(t)\delta \omega_{-}(t)dt
    \label{eqn:total} 
\end{equation}

The function $f_{+,-}(t)$ is determined by the positioning of $\pi$ pulses in each decoupling sequence, given as:

\[
  f_{\text{Ramsey}}(t) =
  \begin{cases}
    1 & \text{if $0<t<\tau$} 
  \end{cases}
\]
\[
  f_{\text{Spin Echo}}(t) =
  \begin{cases}
    1 & \text{if $0<t<\tau/2$} \\
    0 & \text{if $\tau/2<t<\tau$} 
  \end{cases}
\]
\[
  f_{\text{CPN=2}}(t) =
  \begin{cases}
    1 & \text{if $0<t<\tau/4$} \\
    0 & \text{if $\tau/4<t<3\tau/4$}  \\
    1 & \text{if $3\tau/4<t<\tau$}
  \end{cases}
\]

\subsection{Single-Atom Ramsey}
\label{sec:single-atom-ramsey}

Solving Eq.~\eqref{eqn:total} for the Ramsey sequence, we find that the phase acquired will have a dependence on the variance in position and velocity of the atom. However, for our experiments their contribution to the total dephasing is small and we find that the single-atom Ramsey decay time is limited by shot-to-shot fluctuations of the $\delta \omega_0$ term, caused by electric field fluctuations. Therefore, $T_2^*$ is given by the variance of the detuning $\sigma^2_{\delta \omega_0}$:
\begin{equation}
    T_2^* \approx \sqrt{\frac{2}{\sigma^2_{\delta \omega_0}}}
    \label{eq:t2star_to_sigma}
\end{equation}

From the stability measurement of the spectral shifts in Fig.~\ref{fig:line_stability} we find that the fluctuations have a Gaussian distribution. We assume that both the charge on the device $q$ and the background electric field $\mathbf{E}$ fluctuate with variances $\sigma_{q}$ and $\sigma_{\mathbf{E}}$ respectively. To first order, the total variance can be written as:
\begin{equation}
    \sigma^2_{\delta \omega_0}  = \left(\frac{d\delta \omega}{d\mathbf{E}}\right) ^2\sigma^2_{\mathbf{E}} +\left(\frac{d\delta \omega}{dq}\right)^2\sigma^2_{q}
\end{equation}
Using Eq.\eqref{eq:delta_omega}, this expression becomes:
\begin{equation}
\begin{split}
    \sigma^2_{\delta \omega_0}  = 4\pi^2\alpha^2 ((&\frac{k^4}{|\mathbf{r}|^8}q^2_0+\frac{2k^3}{|\mathbf{r}|^6}E_x q_0 + \frac{k^2}{|\mathbf{r}|^4}E_x^2)\sigma_q^2 + \\
    &\frac{k^2}{|\mathbf{r}|^4}q^2_0\sigma^2_{E_x}+|\mathbf{E}|^2\sigma^2_{|\mathbf{E}|})
\end{split}
\end{equation}
where the $x-$direction is along the axis of movement in the distance-dependent measurements. 

\begin{figure}
     \centering
     \includegraphics[width=0.75\linewidth]{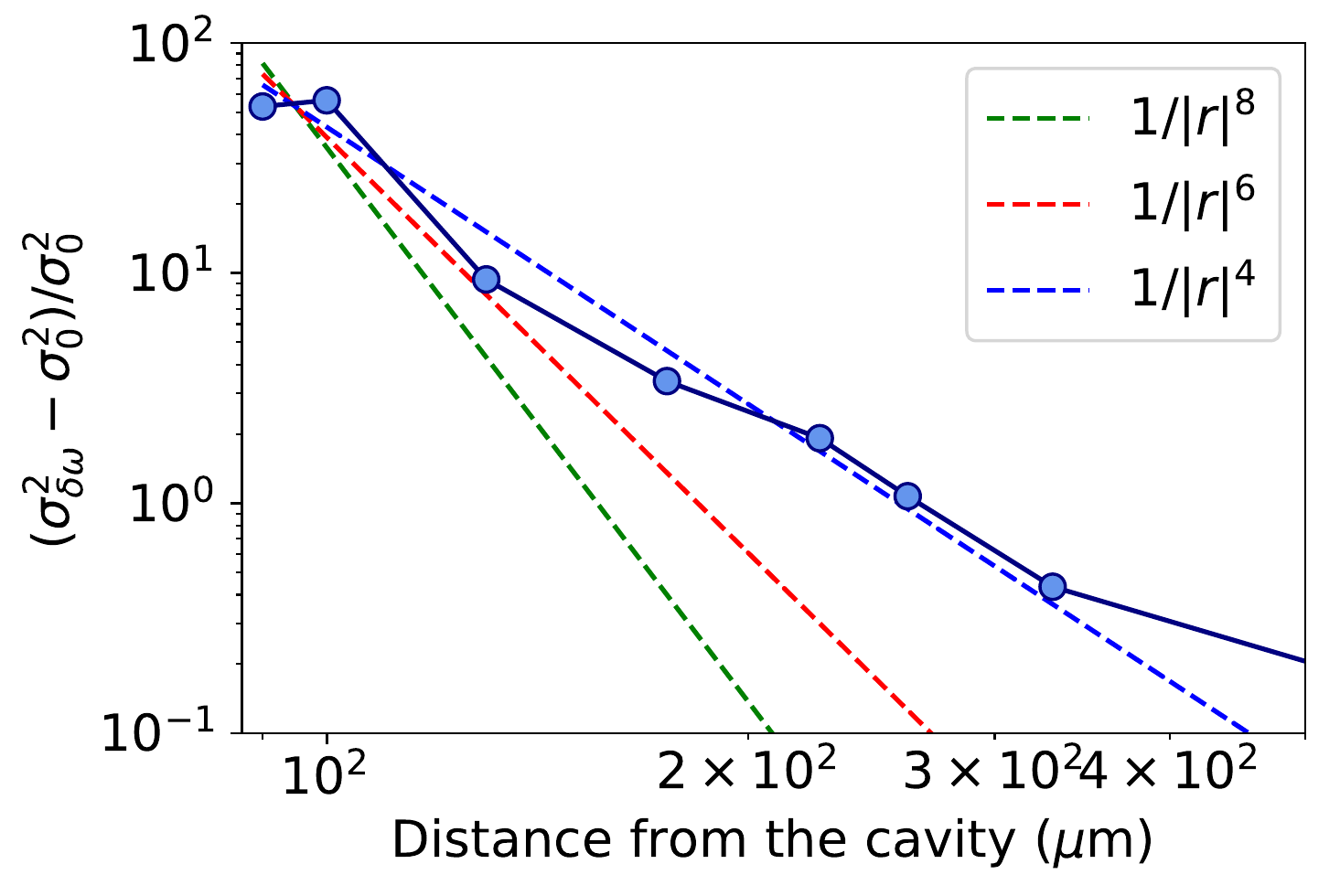}
     \caption{Variance of the Rydberg shift detuning as a function of distance. Plotted are different distance scalings, suggesting that the measured variance scales as $1/|r|^4$. }
\label{fig_varScaling}
\end{figure}

Using this expression for $\sigma^2_{\delta \omega_0}$, we can extract whether the dominant source of noise is charge fluctuations or background electric field fluctuations, due to the difference in the distance scaling of the different terms. We use the measured $T_2^*(r)$ and convert it to variance using Eq.~\eqref{eq:t2star_to_sigma}. We then subtract a constant offset from the variance to account for any contribution to the dephasing which does not scale with distance, such as Doppler dephasing. We use the form $\sigma^2_0=2/(T_2^*(r_{\rm max}))^2$ for this constant offset. This is plotted in Fig.~\ref{fig_varScaling}, showing that the variance $\sigma^2_{\delta \omega_0} - \sigma^2_0$ scales as $|r|^{-4}$. Fitting for the two variances shows that for the distances considered here, the noise is dominated by fluctuations of the electric field background. This model is used to generate the theoretical fit of the Ramsey decay time in Fig. 2(D).

\begin{figure}
     \centering
     \includegraphics[width=0.75\linewidth]{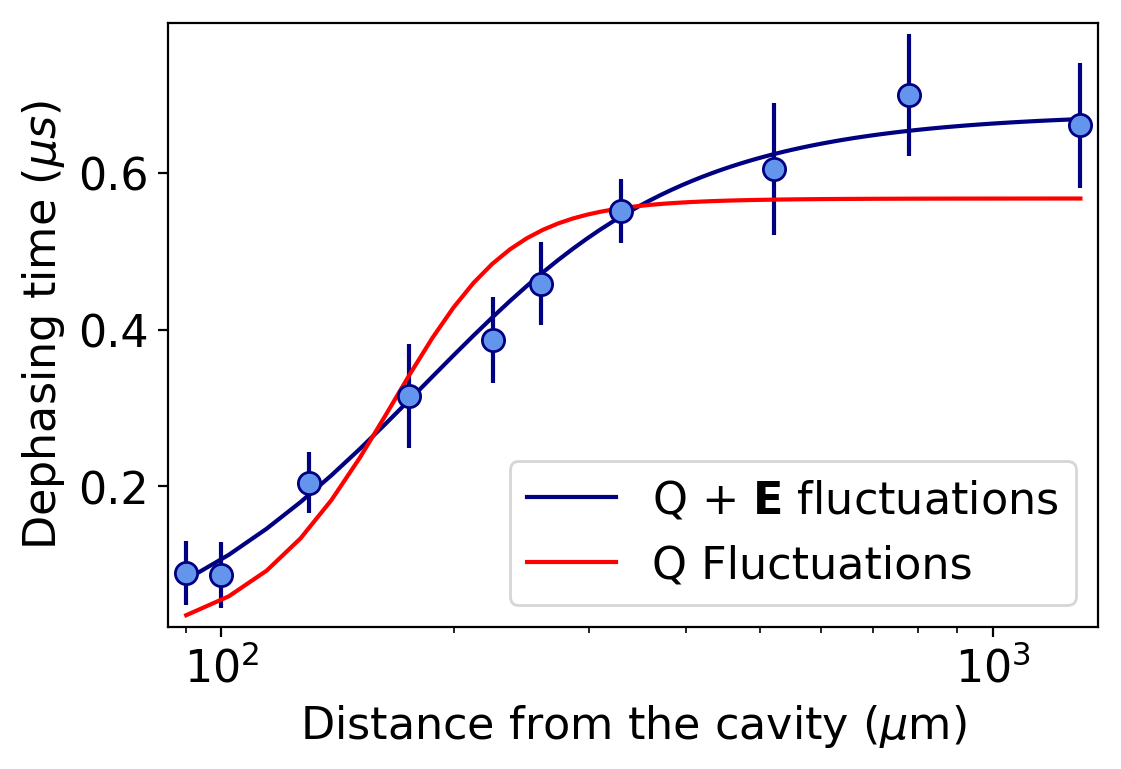}
     \caption{Model comparison of Ramsey decay time. We compare a model taking into account only charge fluctuations and constant baseline $T_2^*$ (red) to the model that incorporates both charge and $\mathbf{E}$ fluctuations (blue).}
\label{fig_ramComparison}
\end{figure}

In Fig.~\ref{fig_ramComparison} we compare a fit with only charge fluctuations and a fit which includes both charge and background field fluctuations to further emphasize the presence of background field fluctuation.

\subsection{Single-Atom Spin Echo}

Solving Eq.~\eqref{eqn:total} for the spin echo pulse sequence we find that the dephasing is insensitive to low frequency noise, such as shot-to-shot fluctuations of $\delta \omega_0$. However, we find that in an electric field gradient, even in the absence of charge fluctuation, the spin echo dephasing rate is limited by thermal effects. The total phase acquired during the spin echo sequence will be given by:
\begin{equation}
    \phi = \mathbf{\nabla} \delta \omega \cdot \frac{\mathbf{v}}{4}\tau^2 
\end{equation}
In our experiment we find that the velocity fluctuations dominate the dephasing, resulting in a phase variance of:
\begin{equation}
    \sigma^2_{\phi} =|\nabla \delta \omega|^2 \frac{\tau^4}{16}\frac{k_b T}{m}
\end{equation}
From this we find that the variance increases as $\tau^4$. This results in an expected decay going as $e^{-(t/T_2)^4}$ as opposed to the typical Gaussian decay. We can then solve for the thermally limited coherence time, $T_{thermal}$:	
\begin{equation}	
    T_{\rm thermal} =2\sqrt{ \frac{\sqrt{2}}{|\nabla \delta \omega|\sqrt{\frac{k_b T}{m}}}}	
\end{equation}	
From this equation we find a distance dependence from the $|\nabla \delta \omega|$ term. By simplifying this equation, we find an approximate scaling of:
\begin{equation}	
    T_{\rm thermal} \propto \frac{r^{2.5}}{q\sqrt{\alpha \sqrt{T}}}	
\end{equation}
We then assume some baseline dephasing of $\sigma_0$ to yield an overall $T_2$ of:
\begin{equation}
    T_2 = \sqrt{\frac{2}{\sigma^2_{0}+\frac{1}{2}|\nabla \delta \omega|\sqrt{\frac{k_b T}{2m}}}}
\end{equation}

This model is used to generate the theoretical fit of the spin echo decay time in Fig. 2(D). Due to the dependence on the electric field gradient, this model can also provide a measurement of the background electric field in not only the $x-$direction, but also orthogonal directions. From this model we extract that there is a background field orthogonal to the $x-$field with a magnitude of \SI{0.4(1)}{\V/\cm}. 

\subsection{Carr-Purcell N=2}
Using Eq.~\eqref{eqn:total} to solve for the two-pulse Carr-Purcell sequence, we find that the atom becomes insensitive to thermal sampling. However, there is a term that is sensitive the the electric field gradient given by:

\begin{equation}
    \phi =\nabla \delta \omega \cdot \frac{\mathbf{a}}{384}\tau^3
\end{equation}
We can rewrite the phase acquired during the CP N=2 pulse sequence when taking into account Eq.\eqref{eq:acc} as:
\begin{equation}
    \phi =\frac{\hbar\tau^3}{384m} |\nabla \delta \omega|^2
\end{equation}
Charge instability will cause variance in $ |\nabla \delta \omega|$, leading to variance in the phase. However, while charge instability sets a limit on the coherence time for the CP N=2, there exists a class of pulse sequences that can decouple dephasing of this form. 

\subsection{Cancelling polynomial phase}
The noise that we have analyzed in this section is unique as it is noise that has a polynomial time dependence. For example, this noise is constant for positional dephasing, linear with velocity dephasing, and quadratic in dephasing due to acceleration. If we assume that the phase accumulated with time has a functional form of:
\begin{equation}
    \phi(t) = \sum_{j=1:k} a_j t^j
\end{equation}
where $a_j$ are constants that may be varying due to some external noise, and $k$ is the maximal polynomial order. We want to design a pulse sequence which will cancel out the accumulated phase $\phi(t)$ regardless of the values of $a_j$. 

We apply $N$ $\pi$ pulses at times $t_2, t_3,...,t_{N+1}$. As a convention, $t_1 = 0$ is the beginning of the phase accumulation and the sequence length is $T$, such that $t_{N+2} = T$. We can now rewrite the accumulated phase as:
\begin{equation}
    \phi(t) = \sum_{i=1:N+1} -(-1)^i \sum_{j=1:k} a_j (t_{i+1}^j - t_{i}^j)
\end{equation}
In the simplest case, $\phi(t) = at$ is the constant rate of phase accumulation. Then:
\begin{equation}
    \phi(t) = \sum_{i=1:N+1} -(-1)^i a (t_{i+1} - t_{i})
\end{equation}
To find the times when we should apply the pulses, we set $\phi(t)=0$ and solve for $t_2,..t_{N+1}$. In this case, we know that one pulse will echo out the phase, so $N=1$:
\begin{equation}
     \phi(t) = a (t_{2} - t_{1}) - a (t_{3} - t_{2}) = 0 
\end{equation}
Furthermore, $t_1 = 0$ and $t_{N+2} = t_3 = T$, so
$$
    a t_{2} - a T + a t_{2} = 0 
$$
\begin{equation}
    t_2 = T/2
\end{equation}

This results in the pulse sequence that we know as the Spin echo, which decouples from DC noise.

For $k>1$, we posit that $N=k$. We will then get a system of $k$ equations with $k$ unknown times. We can split them into contributions for different $a_j$, and set all contributions to zero:
\begin{equation}
    \sum_{i=1:N+1} -(-1)^i (t_{i+1}^j - t_{i}^j) = 0 , \forall j \in(1,N)
\end{equation}

\begin{figure}[h!]
\center
\includegraphics[width=0.75\columnwidth]{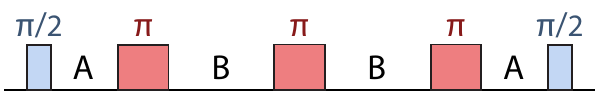}
\caption{Pulse sequence that cancels polynomial phase noise up to $t^2$ contribution.}
\label{fig:poly_pulses}
\end{figure}

When solving these equations for k=N=2 we recover the Carr-Purcell N=2 sequence. Now, we can solve these equations for $k=N=3$:
$$
t_2-t_3+t_4-T/2=0
$$
$$
t_2^2-t_3^2+t_4^2-T^2/2=0
$$
$$
t_2^3-t_3^3+t_4^3-T^3/2=0
$$
and as always, $t_1=0$, $t_5=T$. We would apply the pulses at $t_2,t_3,t_4$.
The equations above in general have multiple solutions, but we restrict the solutions to real numbers, and $T>t_4>t_3>t_2>t_1$. This imposes a unique solution to the set of the equations above, resulting $\pi-$pulses at times:

$$
t_2 = (\frac{2-\sqrt{2}}{4})T 
$$
$$
t_3 = T/2
$$
$$
t_4 = (\frac{2+\sqrt{2}}{4})T 
$$
From this solution we find that the pulse sequence would take the form illustrated in Fig.~\ref{fig:poly_pulses} with evolution periods $A=\left(\frac{2-\sqrt{2}}{4}\right)$ and $B=\frac{\sqrt{2}}{4}$. 

The resulting function is given as:

\[
  f_{\text{k=N=3}}(t) =
  \begin{cases}
    1 & \text{if $0<t<(\frac{2-\sqrt{2}}{4})\tau$} \\
    0 & \text{if $(\frac{2-\sqrt{2}}{4})\tau<t<\tau/2$}  \\
    1 & \text{if $\tau/2<t<(\frac{2+\sqrt{2}}{4})\tau$}\\
    0 & \text{if $(\frac{2+\sqrt{2}}{4})\tau<t<\tau$}
  \end{cases}
\]

\section{Two-atom coherence}

\subsection{W-state Fidelity measurement}
\label{sec:extracting_fidelity}
In order to characterize the fidelity of preparing the state $\ket{W} = \frac{1}{\sqrt{2}} \left( \ket{gr} + \ket{rg}\right)$, we follow the procedure in \cite{levine18}. We can express the fidelity of the entangled state with elements of the density matrix $\rho$:
\begin{align}
\mathcal{F} &= \langle W|\rho|W\rangle  \nonumber \\
&= \frac{1}{2}\left( \rho_{gr,gr}+ \rho_{rg,rg}+ \rho_{gr,rg} + \rho_{rg,gr} \right)
\label{eq:fidelity}   
\end{align}
We extract the diagonal elements $\rho_{rg, rg}$ and $\rho_{gr, gr}$ by measuring the populations, $P_{rg}$ and $P_{gr}$, after preparing the $\ket{W}$ state with a $\pi-$pulse from $\ket{gg}$. 

The off-diagonal elements can be extracted by applying a local phase shift $\Delta \phi = \delta t$ to one of the atoms, embedded in a Ramsey-type sequence: $\pi - \delta t - \pi$. The accumulation of differential phase rotates between $\ket{W}$ and $\ket{D}=\frac{1}{\sqrt{2}} \left( \ket{gr} - \ket{rg}\right)$:
\begin{align}
      \ket{\Psi} &= \frac{1}{\sqrt{2}} \left( \ket{gr}+ e^{i \delta t} \ket{rg} \right)\\
      &= \cos(\delta t/2)\ket{W}+\sin(\delta t/2)\ket{D}  
\end{align}
The final $\pi-$pulse maps the population in $\ket{W}$ to population in $\ket{gg}$. Since the dark state $\ket{D}$ does not couple to $\ket{gg}$, this sequence measures the time evolution of the population in $\ket{W}$. 
For a general density matrix, the amplitude of these oscillations will be equal to the off-diagonal density matrix elements, $\rho_{rg,gr}=\rho_{gr,rg}$ \cite{levine18}: 
\begin{equation}
    P_{gg}(t) = |\rho_{gr,rg}|\cos(\delta t+\phi) + C
\label{eq:coherence_oscillation}
\end{equation}

\begin{figure}[t]
\center
\includegraphics[width=0.9\columnwidth]{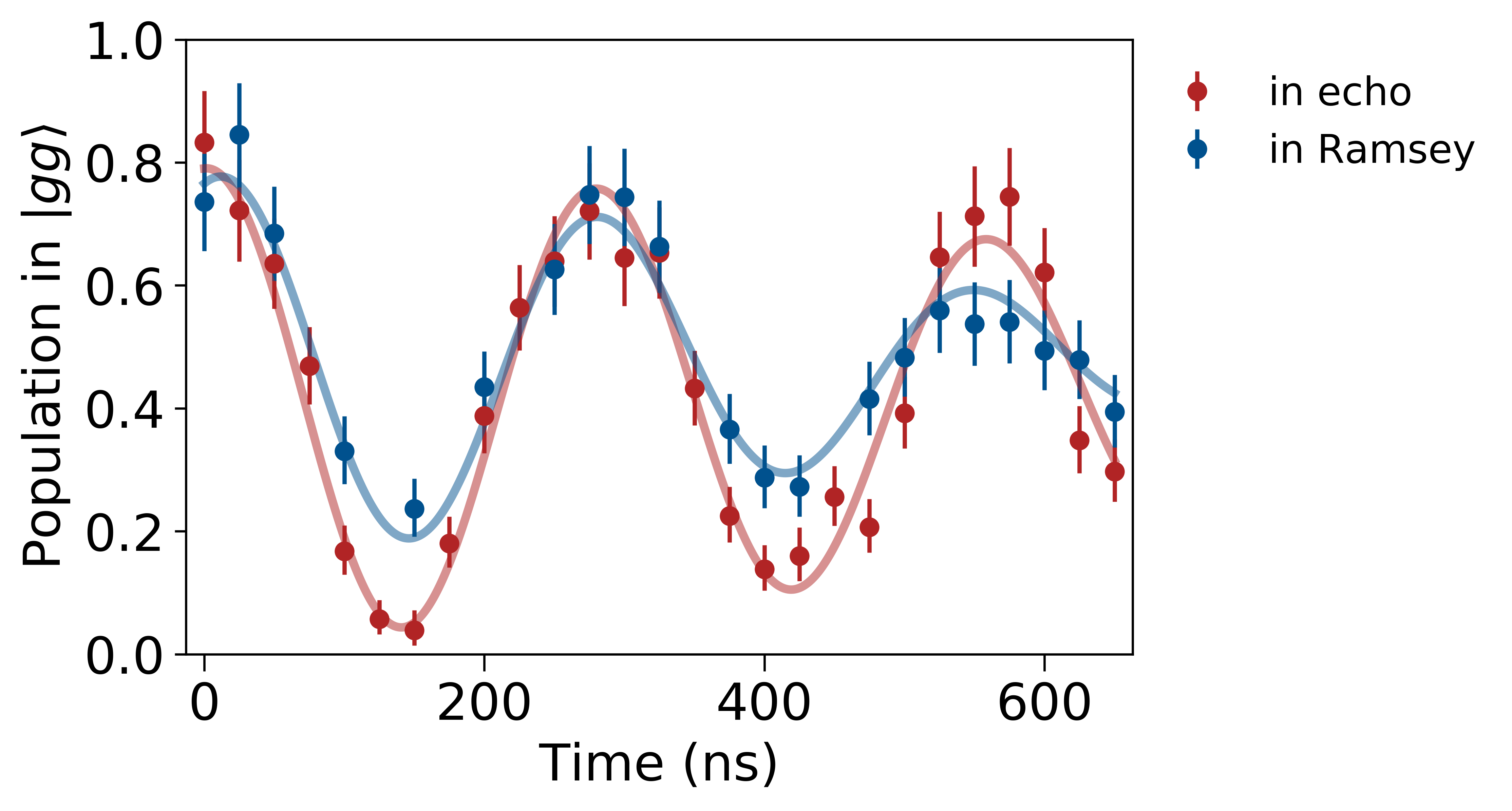}
\caption{$\ket{W}$-state coherence measurement embedded in a Ramsey-type sequence (blue) $\pi - \delta \cdot t - \pi$ and embedded in an echo-type sequence (red) $\pi - \delta \cdot t - 2\pi - t - \pi$. The fitted decays are $\tau_{\rm in\,\,Ramsey} = 570 \pm 10 $ ns, $\tau_{\rm in\,\,echo} = 930 \pm 10 $ ns. }
\label{fig:coherence_in_echo_ramsey}
\end{figure}

In order to mitigate sensitivity to dephasing we embedd this phase shift in a spin echo sequence \cite{levine18} where each period of phase acquisition is fixed to $\SI{1}{\mu s}$. The trap is applied for a varying amount of time within that $\SI{1}{\mu s}$. Fig.~\ref{fig:coherence_in_echo_ramsey} compares two measurements with and without an added echo pulse. For the data in this paper, we use the echo sequence to extract the coherences.

The contrast of the coherence oscillation can either be obtained by taking the difference between the maximum and the minimum points or by using a decaying sinusoidal fit. We use the former because it is independent of the functional form of the decay.

\subsection{W-state fidelity at two distances}
\begin{figure}[h]
\center
\includegraphics[width=0.9\columnwidth]{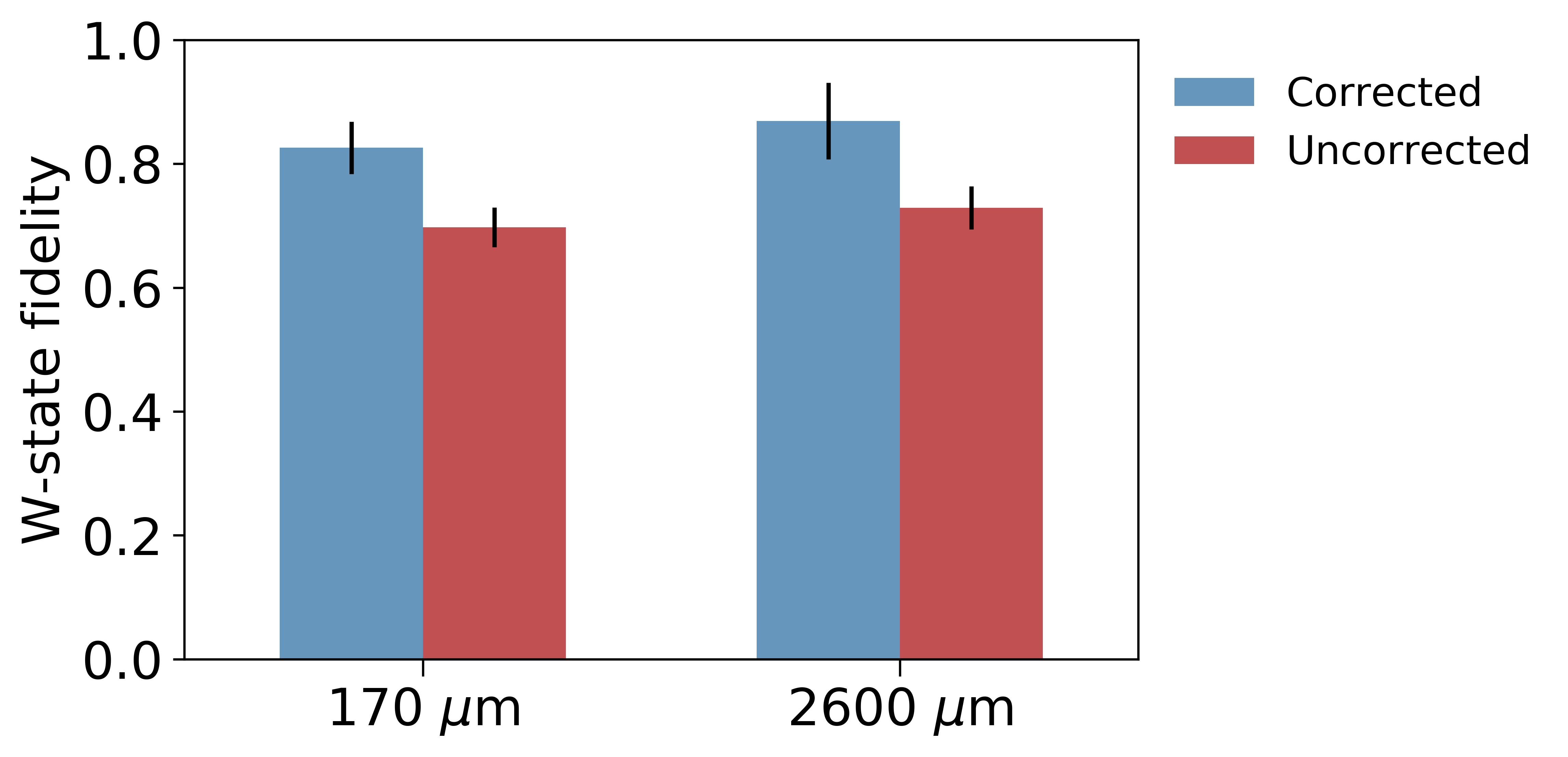}
\caption{Preparation fidelity of the $\ket{W}$-state at two distances from the device. Both the SPAM corrected (blue) and the uncorrected (red) fidelities are shown. We observe little dependence of the fidelity on distance from the device.}
\label{fig:fidelity_near_far}
\end{figure}

We measure the $\ket{W}$-state preparation fidelity at two distances from the device: both far (2.6 mm) and close (170 $\mu$m). We observe little dependence of the fidelity on distance as shown in Fig.~\ref{fig:fidelity_near_far}, suggesting that the preparation fidelity in our case is limited by factors other than electric field noise, for example, laser intensity and phase noise.

\subsection{W-state fidelity at different orientations}
\begin{figure}[h]
\center
\includegraphics[width=0.9\columnwidth]{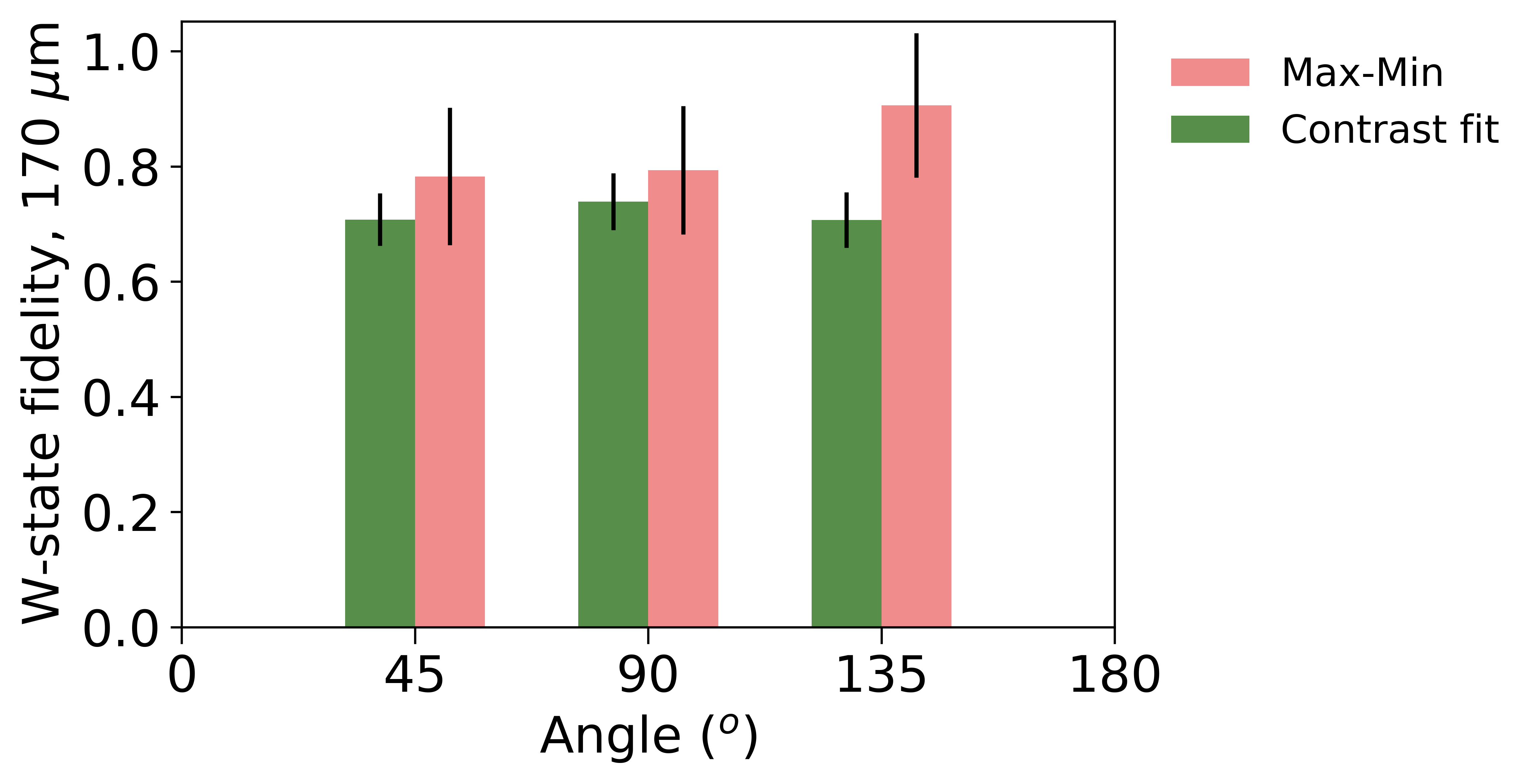}
\caption{$\ket{W}$ state fidelity for three different orientations with inter-atomic distance \SI{3.15}{\mu\m}. The energy differences between the atoms for $\theta = 45^o, 90^o, 135^o$ are \SI{0.11}{\mega\Hz}, \SI{0.63}{\mega\Hz}, \SI{0.84}{\mega\Hz} respectively.}
\label{fig:fidelity_vs_orientation}
\end{figure}

At 170 $\mu$m we explore the $\ket{W}$-state fidelity at three different orientations of the two atoms which are compared in Fig.~\ref{fig:fidelity_vs_orientation}. In each case, the fidelity is calculated both by using the coherence contrast obtained by taking the difference between the maximum and the minimum, as in the rest of the paper, and by using a decaying sinusoidal fit of the form $f(t) = a \cos(\delta t + \phi) e^{-\left( t/\tau \right)^2}$. While the two methods give similar results, the contrast fit has smaller errorbars since the data did not average for as long as data used in the main text. The coherence extracted from the fit gives a stronger indication that the fidelity value is independent of the angle. Fig.~\ref{fig:fidelity_vs_orientation} shows that the $\ket{W}$-state preparation fidelity is insensitive to the electric field gradient between the positions of the two atoms for the values of the gradient at this distance.

\subsection{W-state lifetime }
Similarly to the discussion of the single-atom coherence, here we consider several possible dephasing mechanisms for two atoms in an electric field gradient. However, for two entangled atoms, we need to consider only effects which cause a differential phase accumulation between $\ket{gr}$ and $\ket{rg}$, i.e. processes which rotates $\ket{W}$ into $\ket{D}$:
\begin{equation}
    \ket{\Psi} =  \frac{1}{\sqrt{2}} \left( \ket{gr}+ e^{i \Delta \phi_{rg,gr}} \ket{rg} \right)\\
\end{equation}
where $\Delta \phi_{rg,gr}$ is the differential phase. As we outline below, the main sources of decoherence in our setup are Doppler dephasing and thermal sampling of an electric field gradient. 

We use the Ramsey-type sequence described above $\pi - t -\pi$ and measure the decay time of the population $P_{gg}(t)$:
\begin{equation}
    P_{gg} = \frac{1}{2} \left( 1+ \cos (\Delta \phi_{rg,gr} t) \right)
\end{equation}

When an atom with velocity $\mathbf{v}$ is excited to the Rydberg state, it experiences a Doppler shift $\mathbf{k}_{\rm eff} \mathbf{v}$, where in our case $\mathbf{k}_{\text{eff}} =\mathbf{k}_{420}+\mathbf{k}_{1013}$ for co-propagating beams, which includes the effective wavevectors of the \SI{420}{\nm} and \SI{1013}{\nm} lasers. If the atomic velocity follows a thermal Maxwell-Boltzmann distribution with width $\sigma_v$, this will cause dephasing of the $\ket{W}$ state because each atom in $\ket{r}$ will experience a different velocity sampled from this distribution. 

 Similarly, in the presence of an electric field gradient, each thermal atom will locally sample a different electric field and will experience a different Rydberg spectral shift as a function of its initial position and velocity. The atom samples a different velocity and position for each realization of the experiment, resulting in a quasi-static phase noise.

The differential phase between $\ket{rg}$ and $\ket{gr}$ can be written in terms of the differences between each atom's velocity $\mathbf{v}_{a,b}$ and position $\mathbf{r}_{a,b}$:

\begin{equation}
    \Delta\phi(t) =\nabla \delta \omega \cdot (\mathbf{\Delta r} + \mathbf{\Delta v}t)
\end{equation}
where $\mathbf{\Delta v}=\mathbf{v}_{a}-\mathbf{v}_{b}$ and $\mathbf{\Delta r}=\mathbf{r}_{a}-\mathbf{r}_{b}$. 
After evolving for time $\tau$ the state will become:

\begin{equation}
\ket{\Psi}=\ket{rg}+e^{i(\mathbf{k_{eff} \cdot \mathbf{\Delta v})}+i\int_{0}^\tau \Delta \phi dt 
}\ket{gr}
\end{equation}

The rotation between $\ket{W}$ and $\ket{D}$ caused by the gradient follows an oscillation of the form:
\begin{equation}
    \frac{1}{2}+\frac{1}{2}\cos\Bigl((\Delta\mathbf{r}\cdot\nabla\delta\omega)\tau\Bigl)
\end{equation}
and experiences an overall exponential decay of the form:
\begin{equation}
    e^{-\frac{1}{2}\left(\mathbf{k}\cdot\sigma_{\mathbf{\Delta v}}\right)^2\tau^2 - \frac{1}{2}\left(\nabla\delta\omega\cdot\sigma_{\mathbf{\Delta r}}\right)^2\tau^2 - \frac{1}{8}\left(\nabla\delta\omega\cdot\sigma_{\mathbf{\Delta v}}\right)^2\tau^4}
\end{equation}
where we can neglect the small contribution from the term $\propto \tau^4$. 
Assuming Gaussian distributions with widths $\sigma_{\mathbf{\Delta v}}$ and $\sigma_{\mathbf{\Delta r}}$, we solve for the following expression for the decay time of the $\ket{W}$ state:
\begin{equation}
    T_{W} = \sqrt{\frac{2}{\sigma^2_{\phi}}}
\end{equation}
where
\begin{equation}
\sigma^2_{\phi}=(\mathbf{k}\cdot\sigma_{\mathbf{\Delta v}})^2+\left(\nabla \delta \omega \cdot \sigma_{\mathbf{\Delta r}}\right)^2
\end{equation}
While for larger distances, the Doppler term dominates. For distances around $\SI{130}{\mu\m}$, the contributions of the two terms start to become comparable. A curve generated from this model using only the atom temperature, trap parameters, point-charge field and electric field background is plotted in Fig. 3(D) of the main text.

\subsection{W-State Spin Echo}

For the W-state, we are also able to apply a spin echo sequence to make the coherence insensitive to DC noise in the differential phase acquired. As a result, we find that the coherence time should no longer be limited by the Doppler dephasing and only by the thermal sampling of the electric field gradient. 

\begin{equation}
    \Delta\phi(t) =\nabla \delta \omega \cdot \frac{\mathbf{\Delta v} t^2}{4}
\end{equation}

Similar to the single-atom spin echo, we find that the coherence time of the W-state is limited by the thermal variance of velocities. The variance of the phase acquired during the pulse sequence will be set by the difference between the two atom's velocities, making the decoherence time a factor of $\sqrt{2}$ smaller than the single-atom decay:

\begin{equation}
    T_{2} =2\sqrt{ \frac{1}{|\nabla \delta \omega|\sqrt{\frac{k_b T}{m}}}}
\end{equation}

\begin{figure}[h]
\center
\includegraphics[width=0.75\columnwidth]{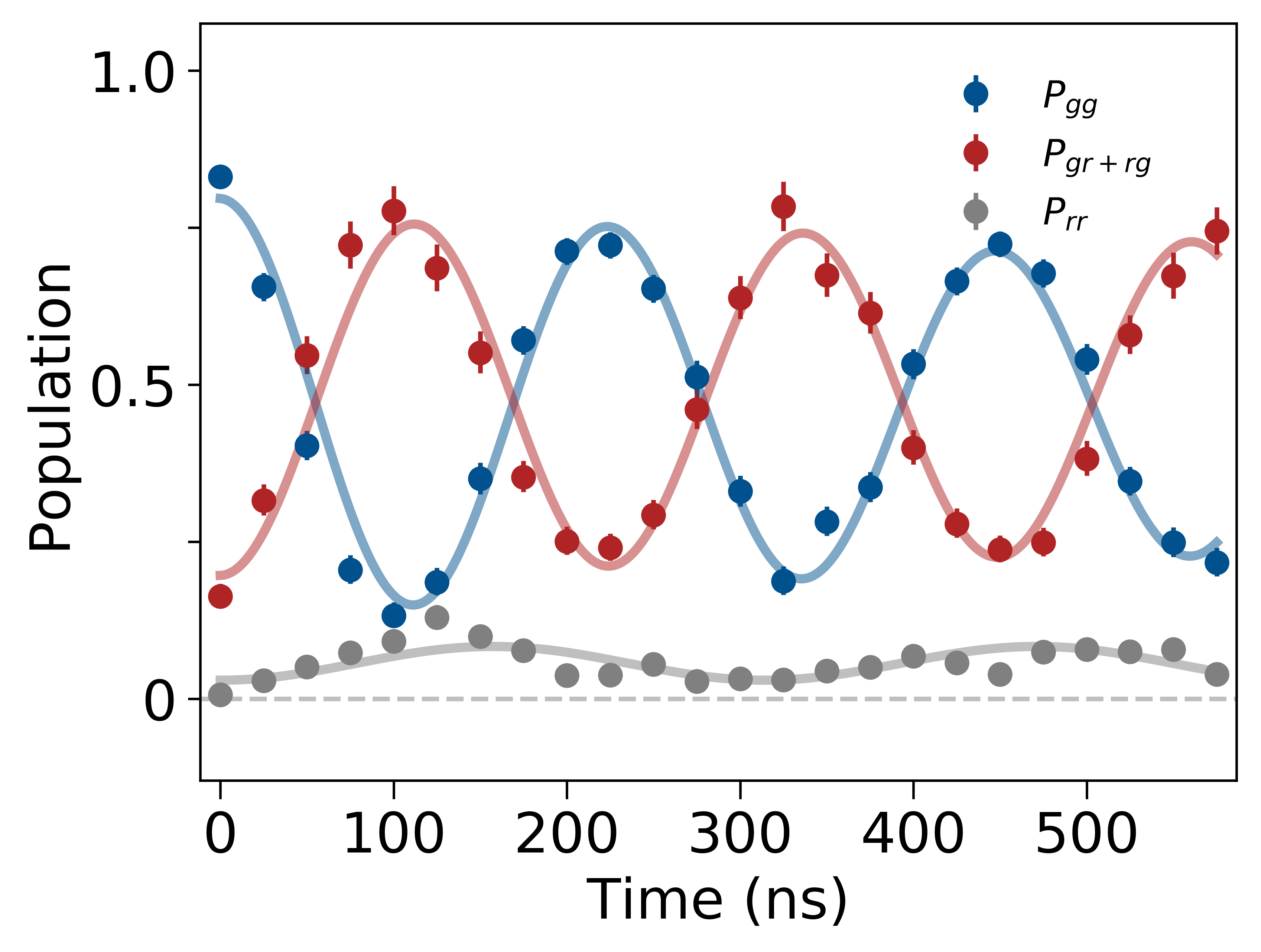}
\caption{Uncorrected Rydberg-blockaded Rabi Oscillation. This data corresponds to the corrected data for the diagonal elements shown in Fig. 3(B) of the main text. We observe a blockaded Rabi frequency of $\Omega_{Blockade}=2\pi\times\SI{4.46(3)}{\MHz}$ between the $P_{\ket{gg}}$ (red) and $P_{\ket{gr}} + P_{\ket{rg}}$ (blue) populations. Additionally, the $P_{\ket{rr}}$ population oscillates at $\Omega=2\pi\times\SI{3.2(1)}{\MHz}$ agreeing with the expected non-blockaded Rabi frequency and therefore expected behavior given uncorrected single-atom loss. }
\label{fig:uncorrected_rabi}
\end{figure}
\subsection{SPAM correction}
We correct our data using the method described in the supplement of \cite{Madjarov-Endres2020}. The main source of preparation error is the loss of an atom after the experimental run is triggered but before the Rydberg excitation occurs. The survival rate of an atom in each tweezer is approximately \SI{90}{\%}. Due to the loss, the uncorrected data has extra contributions from initial states which have only one atom instead of two. Given that we detect Rydberg population through loss, uncorrected data shows increased Rydberg population $P_{\ket{rr}}$. This is apparent in the uncorrected two-atom blockaded Rabi oscillation in Fig.~\ref{fig:uncorrected_rabi}, where the Rydberg population $P_{\ket{rr}}$ is clearly oscillating. In fact, the frequency of this oscillation occurs at the single-atom Rabi frequency and therefore can be explained by single-atom evolution when the dynamics began with only one populated tweezer. To appropriately account for single-atom evolution in a two-atom experiment with finite preparation loss, the SPAM correction requires two extra datasets to be taken concurrently using the same parameters as the data intended for correction. The two extra datasets use the same experimental sequence but each has only one atom present. Additionally, we correct for the \SI{97(1)}{\%} finite probability of initializing the atoms in the correct initial hyperfine ground state, although this results in a much smaller correction.

\section{Electric field gradient}
\label{sec:gradient}
More information can be gained about the electric field configuration and its dependence on UV power by using two atoms. As illustrated in Fig. 4 in the main text, by preparing the W-state and measuring its time evolution, we can detect an electric field gradient at the position of the atoms. 

\subsection{Fitting the oscillation frequency}
By changing the relative orientation of the two atoms, we can study the spatial dependence of the electric field. 
We pick seven different orientations of the two atoms spanning a semi-circle. Our full setup is described in \cite{samutpraphoot20} and here we give a brief overview. We control the positions of the atoms using two galvo mirrors in the path of the optical tweezers. The coordinates of the two galvo mirrors can each be referenced to the nanoscale device. This is done by sending light through the device via the fiber and collecting the fluorescence of a few points on the device on two APDs placed along the tweezer paths when varying the galvo mirror positions, allowing us to map one coordinate system to the other. This gives us nanoscale precision of the orientation of the two atoms relative to each other of $\sim \SI{50}{\nm}$ \cite{samutpraphoot20}.

By measuring $P_{gg}(t)$ for the pulse sequence $\pi - t - \pi$, we can extract the differential phase accumulation during the free evolution time $t$, which corresponds to the energy difference at the positions of the two atoms. We measure this time evolution at each orientations and do a simultaneous fit to all seven with the fitting function for each distance:
\begin{equation}
    P_{gg,j}(t) = a_j \cos(2\pi f_j t + \phi_j)e^{-(t/\tau)^2} + c
\end{equation}
where each orientation $j$ has independent amplitude $a_j$, frequency $f_j$, and phase $\phi_j$ parameters and the decay time $\tau$ and offset $c$ parameters are common to all. Given the dependence of oscillation frequency on angle at each distance, we are now ready to develop a model for the electric field distribution. 

\subsection{2D Electric field gradient model}
Given our measurements, we can try to reconstruct the background electric field $E(x,y)$ in 2D. Note that we do not have access to information about $E_z$ which is along the tweezer propagation direction. The oscillation frequency between $\ket{W}$ and $\ket{D}$ is a measure of the energy difference between the two atoms, here denoted as $A$ and $B$:
\begin{equation}
\nu_{AB} = |\nu_A-\nu_B|
\end{equation}
At each distance, we also obtain the average spectral shift over all orientations and the two atoms:
\begin{equation}
\Delta \nu = \nu(r) - \nu(\infty)
\end{equation}
Note that one could also consider directly measuring $\nu_{AB}$ from the individual energy shifts $\Delta \nu_A$ and $\Delta \nu_B$ of the two atoms. However, these shifts are sensitive to noise and decoherence, making this measurement extremely challenging. It is the entanglement between the two atoms which allows us to perform a differential measurement for which the majority of these noise sources are common mode, that allows us to obtain $\nu_{AB}$ more precisely. 

To reconstruct the electric field $E(x,y)$, we adopt the same model used for the single-atom data. The nanoscale device acts as a point charge $q$ located in the middle of the device and there is a constant homogeneous electric field background $\mathbf{E_0}$, see Eq.~\eqref{eq:line_shift}. In 2D:
\begin{align}
\nu(x,y) =  & \frac{1}{2}\alpha \left( \frac{kQ}{(x^2+y^2)^{3/2}}x + E_x \right)^2 + \nonumber \\
    & \frac{1}{2}\alpha \left(\frac{kQ}{(x^2+y^2)^{3/2}}y +E_y\right)^2
\label{eq:nuxy}
\end{align}

The energy difference between the two atoms can be expressed in terms of the $x-$ and $y-$derivatives of this function at the position between the two atoms and the distance between the two atoms $r_{AB}$: 
\begin{equation}
\nu_{AB}(r,\theta) = r_{AB} \left(\frac{\partial \nu}{\partial x} \cos(\theta) + \frac{\partial \nu}{\partial y} \sin(\theta) \right)
\label{eq:gradient1}
\end{equation}
with $\theta$ as defined in Fig. 4. Assuming the charge is at position $(x_0,y_0)$ with respect to the origin (fiber-device junction depicted in Fig.~\ref{fig:linecharge}(a)), and given that we can move the device along the $x-$direction, so $y=0$, the partial derivatives become:
\begin{align}
\frac{\partial \nu}{\partial x} &=-\alpha kQ \left( \frac{2kQ\tilde{x}}{r^6}+\frac{E_x(2\tilde{x}^2-y_0^2)-3E_y\tilde{x}y_0}{r^5}\right) , \\ \nonumber
  \frac{\partial \nu}{\partial y}&=\alpha kQ \left(\frac{2kQy_0}{r^6} + \frac{E_y(\tilde{x}^2-2y_0^2)+3E_x\tilde{x}y_0}{r^5}\right)
\end{align}
where $\tilde{x}=x-x_0$ and $r = \sqrt{(x-x_0)^2+y_0^2}$. 

To get a sense of the distance-scaling, we consider a charge at the origin ($x_0=y_0=0$). The gradient components further simplify to:
\begin{align}
\frac{\partial \nu}{\partial x} &= -2\alpha kQ \left( \frac{kQ}{r^5} + \frac{E_{x}}{r^3}\right)  \nonumber \\
\frac{\partial \nu}{\partial y} &= \alpha kQ \frac{E_{y}}{r^3}
\label{eq:gradient0}
\end{align}

We can see that the angle of maximum gradient of the energy is given by:
\begin{equation}
    \theta_{max} = \arctan(-\frac{1}{2}\frac{E_{y}}{\frac{kQ}{{r}^2}+E_{x}})
\end{equation}

\begin{figure}[h!]
\center
\begin{overpic}[width=0.75\columnwidth]{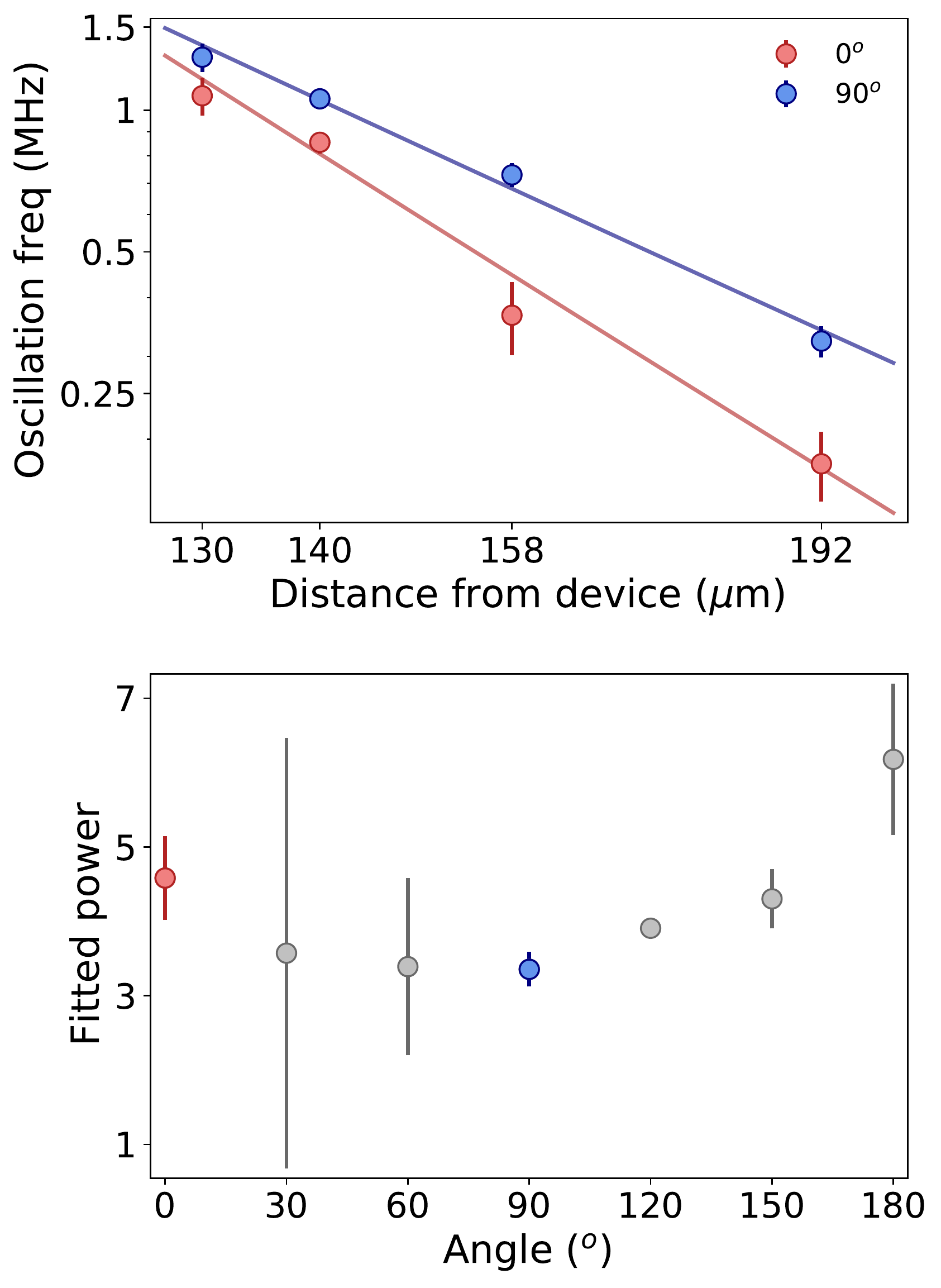}
\put(-5,95){\textbf{(a)}} \put(-5,44){\textbf{(b)}}\end{overpic}
\caption{Gradient scaling with distance from the device. \textbf{(a)} The oscillation frequency at two angular orientations of the atom pair. \textbf{(b)} Extracting the power law scaling $a/r^n$ of the gradient magnitude for each angular orientation measured. For $\theta = 0^o$: $n = 4.6 \pm 0.6$ where we expect $n=5$. For $\theta = 90^o$: $n = 3.4 \pm 0.2$ where we expect $n=3$.}
\label{fig:power_law} 
\end{figure}

Using the fitted oscillation frequencies $f_j(r,\theta)$, we can extract this power law scaling. At each orientation angle $\theta$ we take the four fitted oscillation frequencies and fit a power law of the form:
\begin{equation}
    f_j(x) = a\frac{1}{x^n} 
\end{equation}
The fitted powers are plotted as a function of angle in Fig.~\ref{fig:power_law}. In our convention, $x-$axis lies along $\theta = 0^o$ and the $y-$axis lies along $\theta = 90^o$. We fit the following powers: For $\theta = 0^o$: $n = 4.6 \pm 0.6$ where we expect $n=5$ given Eq.~\eqref{eq:gradient0} since $E_x\approx 0$; For $\theta = 90^o$: $n = 3.4 \pm 0.2$ where we expect $n=3$ for a finite value of $E_y$. Given that we have only four points to fit a power law, these numbers agree with our expectation.

\subsection{Fitting the model to the data}
We can now use this model to reconstruct the electric field $E(x,y)$. We use the expression for the energy difference between the two atoms $\nu_{AB}(x,\theta)$ in Eq.~\eqref{eq:gradient1} to fit the measured oscillation frequencies $f_j(x,\theta)$ as a function of angle $\theta$ and distance to the device $x$. We also use the expression for the Rydberg spectral shift $\nu(r)-\nu (\infty)$ from Eq.~\eqref{eq:nuxy} to fit the measured Rydberg spectral shifts at each distance. We combine the two and obtain $q = \SI{190(10)}{e}$, and $(E_x,E_y) = (-0.02 (1), -0.51(3))\SI{}{\V/\cm}$. Note that if we perform a fit to only the gradient data $\nu_{AB}(x,\theta)$ with Eq.~\eqref{eq:gradient1}, we obtain values which are the same within errorbars: $q = \SI{196(17)}{e}$, and $(E_x,E_y) = (-0.03 (2), -0.47(4))\SI{}{\V/\cm}$.

\begin{figure}[h]
\center
\includegraphics[width = 0.75\columnwidth]{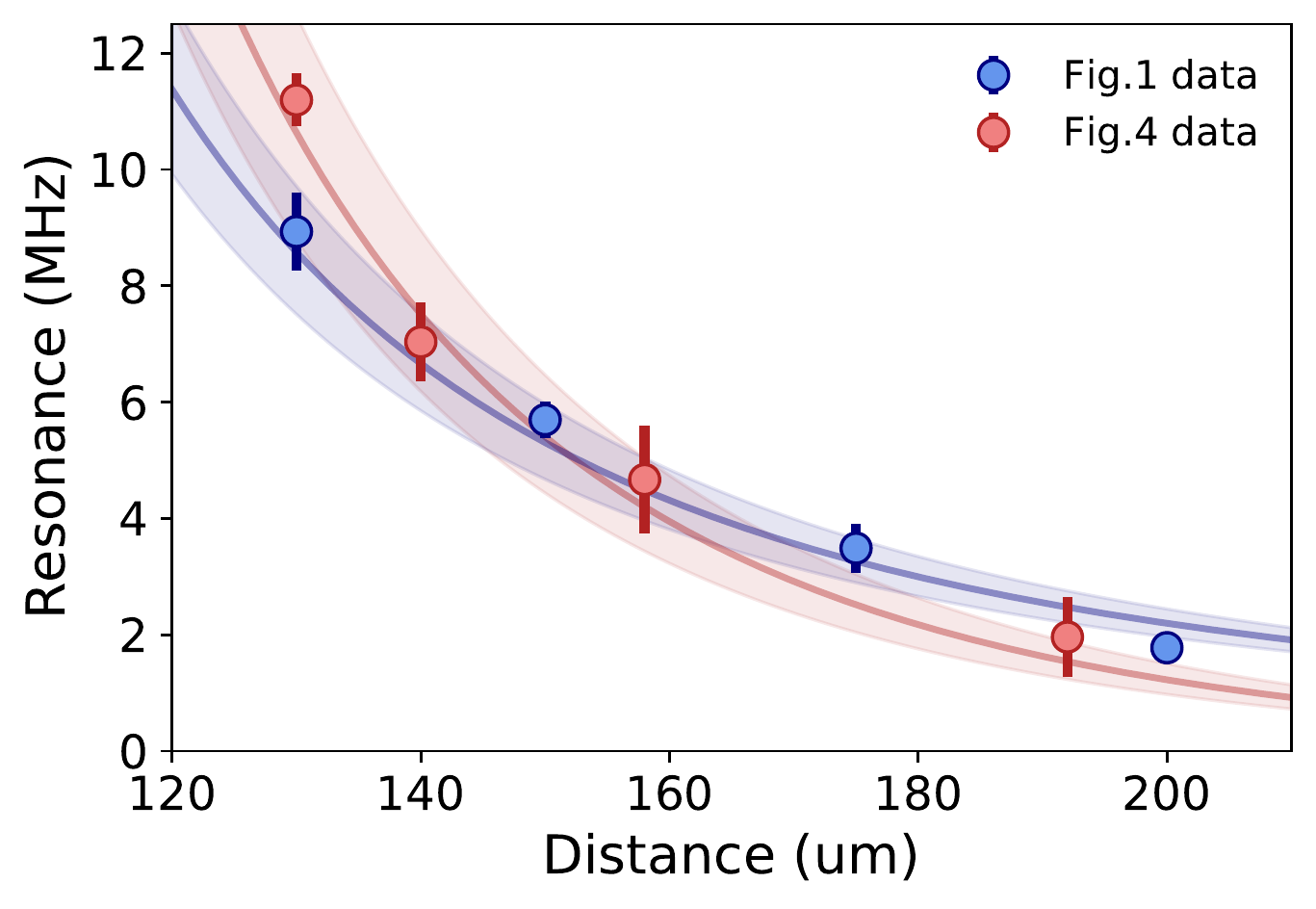}
\caption{Rydberg spectral shift as a function of distance from the device. Comparing data used in Fig. 1 in the main text (blue) where the UV value was optimized for each point, and data from Fig. 4 (pink) where the UV value was kept at the value that minimized the shift at 160$\mu$m for each point. The two datasets were taken 6 months apart but a similar electric field environment could be stabilized.}
\label{fig:fig4_line}
\end{figure}

These values compare well with the data from Fig. 1 in the main text, which finds $q = 126(11)$ and $Ex = 0.049(12)$. In the two-atom data, the UV was kept constant, while in the single-atom data it was varied to minimize the Rydberg spectral shift. The two are plotted together in Fig.~\ref{fig:fig4_line} along with the corresponding model fit values. While different UV powers are necessary to minimize the Rydberg spectral shift over larger distances, over the small distance range of 130-192 $\mu$m used here, the same UV power produces the minimum Rydberg spectral shift, except at the 130 $\mu$m point which begins to deviate.

\subsection{Effect of UV on electric field gradient} 
\label{sec:effect_of_uv_on_gradient}
We use a similar approach to reconstruct the total electric field $(E_x,E_y)$ at a different UV power. We combine the measurement of the energy difference between the two atoms $\nu_{AB}(x,\theta)$ with the shift of the Rydberg spectrum referenced to the infinite-distance energy of the previous measurement using the fitted values $q,E_x,E_y$ from above. We have verified that the UV power shift used does not change the Rydberg spectral shift at $1300 \mu$m (see Fig.~\ref{fig:shift_vs_uv}), which we take to be the infinite-distance value. For the UV-shifted data we obtain 
$q = \SI{201(6)}{e}$, and $(E_x,E_y) = (-0.25(1), -0.528(4))\SI{}{V/cm}$. This results in $\Delta q$ = $11(12)$, and $(\Delta E_x, \Delta E_y) = (-0.23(3),-0.02(3))\SI{}{V/cm}$ relative to the UV-shifted data. This suggests that at $\SI{160}{\mu m}$ a change of UV power of $\sim\SI{200}{nW}$ mostly affects the background field in $x$ rather than the charge on the SiN device. 

Finally, we point out that the reconstruction of the electric field environment relies on the point charge and constant homogeneous background electric field model. Nevertheless, there are other dielectric surfaces present in the vacuum system in the vicinity of the device which could also be affected by changing the power of the UV. A detailed study of these surfaces is beyond the scope of this paper, but their presence can be sensed by the behavior of the Rydberg spectral shift with UV at close and at far distances Fig.~\ref{fig:shift_vs_uv}. At far distances, we have seen that changing the UV power affects the background electric field by at most $\Delta |\mathbf{E}| \leq \SI{0.05}{V/\cm}$ (Section~\ref{subsection:effect_of_uv_1}). This behavior is not captured by our minimal model. The reconstruction of the electric field given the UV-shifted data in Fig. 4 suggests that tuning the UV power over this range does not change the charge on the device but rather changes the background field by 0.23(3) V/cm. Having an additional surface which responds differently to the change of UV power could reconcile this by adding another distance-dependent electric field. Nevertheless, it is clear that tuning the UV power even by $\SI{200}{nW}$ can affect the charges on dielectric surfaces and can have a sizeable effect on the electric field environment in hybrid systems.

\subsection{Anisotropic Rydberg blockade softening}
\begin{figure}[h!]
\center
\begin{overpic}[width = 0.75\columnwidth]{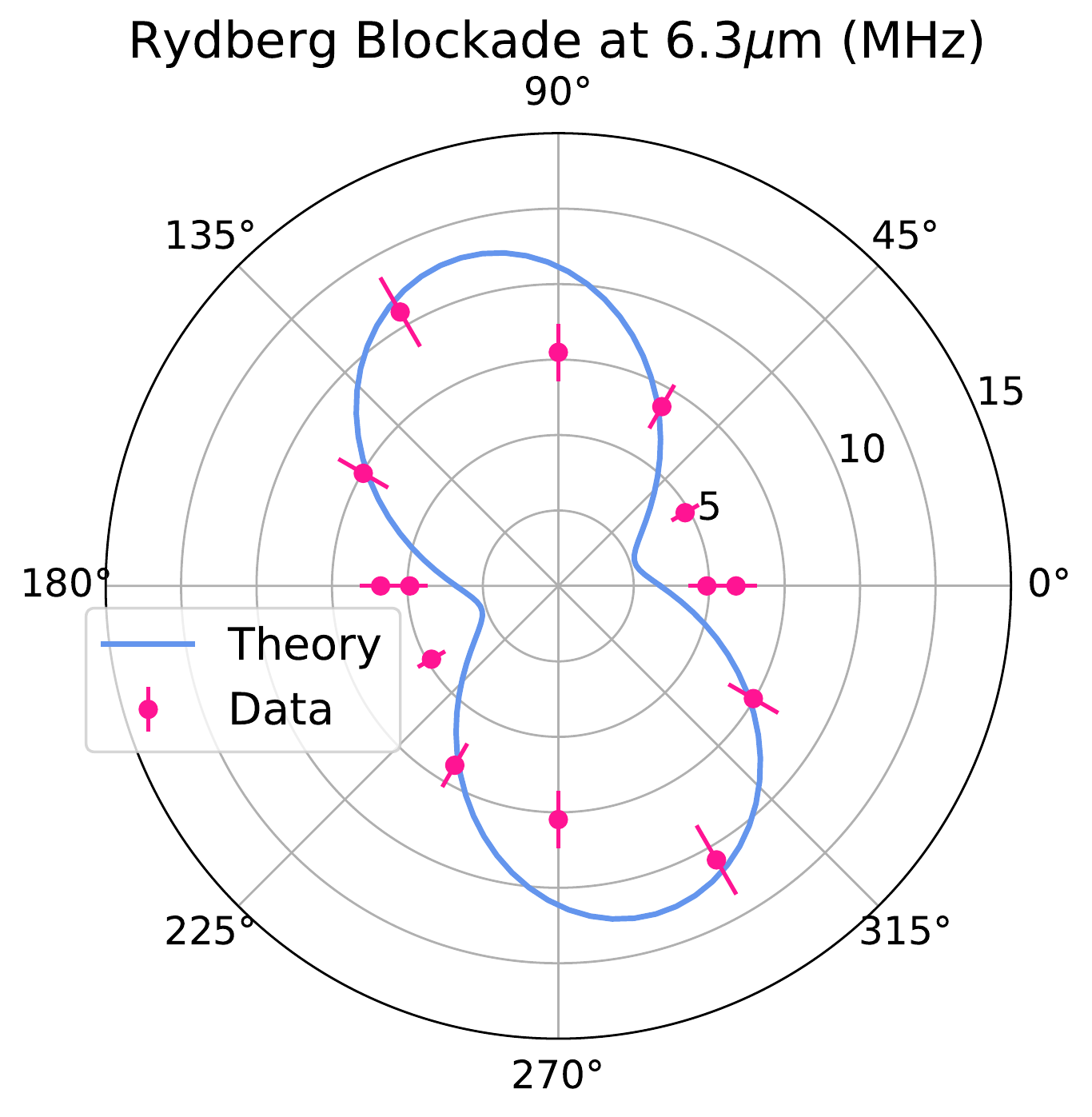}
\end{overpic}
\caption{Rydberg blockade softening at an inter-atomic distance of \SI{6.3}{\um}. Polar plot of measured (pink) and calculated (blue) Rydberg-Rydberg interaction strength. We find reasonable agreement between data and theory when using the electric fields found by fitting the measured gradient as shown in Fig. 4(C) of the main text.}
\label{fig:blockade}
\end{figure}

The presence of an electric field from the device admixes nearby energy levels to the $\ket{70 S}$ state, which are not circularly symmetric. This makes the Rydberg-Rydberg interaction strength anisotropic as a function of orientation of the two atoms. To probe this, we measure Rydberg blockade softening at \SI{140}{\um} from the device and at an inter-atomic distance of \SI{6.3}{\um}, manifested in population in the $\ket{rr}$ state after a $\pi$-pulse:
\begin{equation}
    P_{rr} \approx \left( \frac{\Omega}{U} \right)^2
\end{equation}
Using this equation, we extract the Rydberg-Rydberg interaction $U$ as a function of the relative angle $\theta$ between the inter-atomic axis and the $x-$axis, as depicted in Fig. 4 in the main text. In Fig.~\ref{fig:blockade} we compare the measured values of the Rydberg-Rydberg interaction with numerical calculations using the Pair Interaction Python package~\cite{Weber2017}. In the calculation, we use the fitted background electric fields from the gradient measurements shown in Fig. 4(C) of the main text. We find good agreement between our measured data and theory, as one expects to have more blockade violations in the axis perpendicular to the electric field vector. This independent measurement provides further evidence that the electric field environment sensed through the gradient measurements is correct.

\section{Optimizing the Rydberg quantum number $n$}

In this experiment, we used the $\ket{70S}$ Rydberg state of $^{87}$Rb. 
However, the principal quantum number $n$ could be optimized by taking into account the electric field noise measured in this paper. Since different properties scale differently with $n$, a better choice of $n$ could be made for Rydberg atoms near our device. In particular, the polarizability $\alpha$ scales very rapidly as $\alpha \propto n^7$ and decreasing $n$ would decrease the sensitivity to electric field noise. One could optimize $n$ for two-atom gate fidelity, which when implemented with Rydberg atoms, typically relies on Rydberg blockade with $U \gg \Omega$. Since $U \propto C_6 \propto n^{11}$ and in our experiment $U/\Omega \sim 230$, $n$ could be decreased to $n\sim 55$ without inducing significant Rydberg blockade errors, but reducing the sensitivity to electric field noise by a factor of $\sim 5$. While it is important to note that the lifetime of the Rydberg state decreases with $n$: $\tau_{sp} \propto n^3$ for spontaneous emission and $\tau_{BB}\propto n^2$ for black-body radiation, in the range of $n$ between 50 and 70, this should not limit the gate fidelity.
\clearpage


\bibliography{ryd-bib}

\bibliographystyle{Science}

\end{document}